\documentclass[%
 reprint,
superscriptaddress, 
 amsmath,amssymb,
 pre
]{revtex4-1}

\usepackage[utf8]{inputenc}
\newcommand{\vect}[1]{\boldsymbol{#1}}

\usepackage[T1]{fontenc}
\usepackage{csquotes}

\usepackage{graphicx}
\usepackage{dcolumn}
\usepackage{bm}
\usepackage{float}
\usepackage[dvipsnames]{xcolor}
\colorlet{green1}{green!40!black!100!}

\begin{document}

\preprint{APS/123-QED}

\title{Non-monotonic electrophoretic mobility of rod-like polyelectrolytes by multivalent coions in added salt}

\author{Hossein Vahid}
\affiliation{Department of Applied Physics, Aalto University, P.O. Box 11000, FI-00076 Aalto, Finland}
\affiliation{Department of Chemistry and Materials Science, Aalto University, P.O. Box 16100, FI-00076 Aalto, Finland}
\affiliation{Academy of Finland Center of Excellence in Life-Inspired Hybrid Materials (LIBER), Aalto University, P.O. Box 16100, FI-00076 Aalto, Finland}

\author{Alberto Scacchi}
\affiliation{Department of Applied Physics, Aalto University, P.O. Box 11000, FI-00076 Aalto, Finland}
\affiliation{Department of Chemistry and Materials Science, Aalto University, P.O. Box 16100, FI-00076 Aalto, Finland}
\affiliation{Academy of Finland Center of Excellence in Life-Inspired Hybrid Materials (LIBER), Aalto University, P.O. Box 16100, FI-00076 Aalto, Finland}

\author{Maria Sammalkorpi}
\affiliation{Department of Chemistry and Materials Science, Aalto University, P.O. Box 16100, FI-00076 Aalto, Finland}
\affiliation{Academy of Finland Center of Excellence in Life-Inspired Hybrid Materials (LIBER), Aalto University, P.O. Box 16100, FI-00076 Aalto, Finland}
\affiliation{Department of Bioproducts and Biosystems, Aalto University, P.O. Box 16100, FI-00076 Aalto, Finland}

\author{Tapio Ala-Nissila}
\email{tapio.ala-nissila@aalto.fi}
\affiliation{Department of Applied Physics, Aalto University, P.O. Box 11000, FI-00076 Aalto, Finland}
\affiliation{Quantum Technology Finland Center of Excellence, Department of Applied Physics, Aalto University, P.O. Box 11000, FI-00076 Aalto, Finland}
\affiliation{Interdisciplinary Centre for Mathematical Modelling and Department of Mathematical Sciences, Loughborough University, Loughborough, Leicestershire LE11 3TU, United Kingdom}

\begin{abstract}
{
It is well established that when multivalent counterions or salts are added to a solution of highly-charged polyelectrolytes (PEs), correlation effects can cause charge inversion of the PE, leading to electrophoretic mobility (EM) reversal. In this work, we use coarse-grained molecular dynamics simulations to unravel the less understood effect of coion valency on EM reversal for rigid DNA-like PEs. We find that EM reversal induced by multivalent counterions is suppressed with increasing coion valency in the salt added and eventually vanishes. 
Further, we find that EM is enhanced at fixed low salt concentrations for salts with monovalent counterions when multivalent coions with increasing valency are introduced. However, increasing the salt concentration causes a crossover that leads to EM reversal which is enhanced by increasing coion valency at high salt concentration. 
Remarkably, this multivalent coion-induced EM reversal persists even for low values of PE linear charge densities where multivalent counterions alone cannot induce EM reversal. These results facilitate tuning PE-PE interactions and self-assembly with both coion and counterion valencies.
}
\end{abstract}

\date{\today}


\maketitle

\paragraph*{Introduction.}
Charge reversal (CR) is a rich and ubiquitous phenomenon in chemical and biological processes.
It consists of a charged molecule in solution gaining an overcharge by condensation of the ions of opposite charge and is often observed in strongly correlated systems~\cite{strauss1954}. 
CR is relevant for electrostatic attraction between like-charged rods mediated by multivalent ions~\cite{de1995, grosberg2002}, self-assembly of polyelectrolytes (PEs)~\cite{decher1997}, aggregation of DNAs~\cite{koltover1998, besteman2007, besteman2007-2}, and stabilization of colloids and proteins~\cite{sanchez2001, hotze2010}.
Recent scrutiny of CR include experiments~\cite{besteman2004, besteman2005, zomer2016, wang2018}, theory~\cite{kanduvc2010, kanduvc2011, buyukdagli2020, buyukdagli2021, yang2023}, and computer simulations~\cite{deserno2001, luan2010,tanaka2003, yang2023}.

PEs are polymers with ionizable functional groups that dissociate into charged polymers and their solvated counterions in aqueous solutions.
Properties of the PE's electric double layer, which is intimately connected to CR, are known to depend on the PE charge and ion characteristics, such as size, shape, valency, and concentration.
These factors can be varied to control features such as ion condensation and PE-PE interactions; see, e.g., Refs.~\onlinecite{nguyen2000, becker2012, szilagyi2014, antila2014, antila2015b, Antila2016, antila2017, vahid2022, vahid2023, yang2022}.
Traditionally, multivalent counterions and high line charge density of rod-like macromolecules were believed to be necessary requirements for CR~\cite{de1995, bloomfield1997, Shklovskii1999, Shklovskii1999-2, solis2000, nguyen2000, grosberg2002, tanaka2004, besteman2007}.
However, it has been shown, by means of Monte Carlo and molecular dynamics simulations~\cite{gonzales1985, deserno2001, jimenez2006}, that also large monovalent ions can potentially reverse the PE charge, mainly due to steric interactions. On the other hand, in Ref.~\cite{wang2015} CR of flat surfaces was reported for monovalent counterions and multivalent coions, and attributed to the greater depletion of multivalent coions as compared to the monovalent counterpart close to the surface.
In contrast to this, Refs.~\cite{tanaka2003, antila2017} show that multivalent coions {\it suppress} CR of rod-like PEs induced by multivalent counterions.

A natural way to experimentally probe the existence of CR is to measure the electrophoretic mobility (EM)~\cite{besteman2007}. 
To this end, it is common to measure the drift velocity $\upsilon$ of the molecules in solution induced by a small electric field.
The EM can then be obtained as $\mu \equiv \upsilon/E$, where $E$ is the strength of the applied electric field.
Differences in $\mu$ can be exploited to separate and purify molecules~\cite{de1986, shaffer1989, de1990, kaper2003, danger2007}. Additionally, the ability of PEs to form agglomerates depends on their zeta potential $\zeta$, which can also be estimated via the EM~\cite{nowack2007, obregon2019}.
Hence, a change in the direction of motion of a PE experiencing an external electric field implies a change of sign of the EM and thus its reversal.

According to the Smoluchowski theory of electrophoresis, the EM is proportional to the electrostatic potential at the slipping plane~\cite{von1903}.
The latter is accessible via the ion distributions around the PE. Mean-field theories, such as the Poisson-Boltzmann theory, are among the practical methods to compute ion distributions. 
However, they are approximate and neglect ion correlations~\cite{netz2000,  grosberg2002, chu2007} and finite ion sizes~\cite{borukhov1997, quesada2003, li2009}.
To overcome these limitations, approaches going beyond the weak-coupling Poisson-Boltzmann theory can be employed, which include, e.g., strong-coupling theories~\cite{gronbech1997,rouzina1996, moreira2000, netz2001,netz2003, naji2005, hatlo2010, buyukdagli2012, buyukdagli2014, buyukdagli2014-2, buyukdagli2017}, classical density functional theories~\cite{gonzalez2018, cats2021, cats2022, bultmann2022}, or particle-based simulations~\cite{deserno2000,deserno2001, bagchi2020, vahid2022, vahid2023, yang2022, yang2023}.

Motivated by the complexity of the CR response summarized above, by means of coarse-grained molecular dynamics simulations we systematically investigate the less established effect of coion valency on the effective electrostatic potential and EM of rod-like PEs (such as, e.g., DNA~\cite{gelbart2000}, F-actin~\cite{tang1997, angelini2003}, and tobacco virus~\cite{bernal1941}) under biologically relevant conditions. We fix the size of the hydrated ions and address the condensation response of ions around a DNA-like PE for a wide range of ion valency ratios between counterions and coions at different salt concentrations. We find that multivalent coions can suppress CR induced by multivalent counterions but also induce CR in the presence of monovalent ones. EM reversal is also studied as a function of the PE linear charge density and in different solvent environments via variation of the dielectric constant.

\paragraph*{Model and theory.}
The simulation setup consists of mobile spherical ions in an implicit solvent interacting with a fixed rigid rod-like model of PE.
The PE model is constructed of $120$ spherical coarse-grained beads (force centers) with a charge of $-e$ each, lined up along the $z$ axis in the center of a periodic $20\times20\times20$ nm$^3$ box.
The centers of the PE beads are spaced by $0.167$ nm, which results in a linear charge density of $\lambda_0=-6 \ e/{\rm nm}$, close to the linear charge density of DNA molecules~\cite{kominami2019molecular}.
The PE beads and ions all interact via the Weeks-Chandlers-Andersen~\cite{anderson} potential 
\begin{equation}
    U^{ij}(r)=4\epsilon\left[\left(\frac{\sigma^{ij}}{r}\right)^{12}-\left(\frac{\sigma^{ij}}{r}\right)^{6}+\frac{1}{4}\right],\>\> r\leq r_{\rm c},
\end{equation}
where $r$ is the distance between two interacting particles, $r^{ij}_{\rm c}=2^{1/6}\sigma^{ij}$ the cut-off radius, and the depth of the potential well is set to $\epsilon=0.1$ kcal mol$^{-1}$.
Here, $\sigma^{ij}$ is the interspecies diameter of the interacting pair $i,j$, and is defined as the arithmetic average $\sigma^{ij}=(\sigma^i+\sigma^j$)/2.
We set $\sigma^{\rm PE}=2$ nm and $\sigma^{\rm c}=\sigma^{\rm a}=\sigma=0.5$ nm, where the superscript c indicates cations (counterions) and a anions (coions). 

Additionally, electrostatic interactions are calculated in real space with a cutoff of $1.5$ nm using pairwise Coulombic potential, defined as 
\begin{equation}
\beta e V_{\rm C}(r)= Z^{i} Z^{ j} \ell_{\rm B}/r.    
\end{equation}
Here $Z$ corresponds to the charge valency (hereafter we use the compact notation $Z^{\rm c}$:$Z^{\rm a}$ to denote the valencies of the ions), $\ell_{\rm B}=\beta e^2/(4\pi \varepsilon_0\varepsilon_{\rm r})=0.7$ nm is the Bjerrum length in water, and $\beta=1/k_{\rm B}T$, where $k_{\rm B}$ is the Boltzmann constant, and $T$ the temperature of the system. For the implicit solvent, $\varepsilon_{\rm r}$ and $\varepsilon_{\rm 0}$ express, respectively, the relative dielectric constant of the solvent ($\varepsilon_{\rm r}=78$ for water) and vacuum permittivity. Note that in the last part of the work we systematically vary the value of $\varepsilon_{\rm r}$, consequently also of $\ell_{\rm B}$.
Beyond the cutoff, long-range electrostatic interactions are accounted for using the Particle-Particle Particle-Mesh (P$^3$M) summation procedure~\cite{plimpton1997} with relative force accuracy of $10^{-5}$.

All simulations are carried out using the {\it 3Mar2020} stable version of LAMMPS~\cite{plimpton1995, brown2009, Thomson2022} in the $NVT$ ensemble with the Nos\'e-Hoover thermostat~\cite{nose1984, hoover1985} at $T=300$ K, where a coupling constant of $0.2$ ps has been used. We set the integration time step to $2$ fs, while the trajectories are recorded every ps for a total run of $20$ ns, of which the first $5$ ns are discarded as equilibration time.
The initial configurations are prepared with Moltemplate~\cite{jewett2021}.
We consider ions ranging from monovalent to hexavalent, i.e. $\vert Z^{i}\vert =1-6$, at different salt concentrations $\vert Z^{i}\vert c^i$ up to $1$ M, where $c^{i}$ is the molar concentration of the $i^{\rm th}$ salt species.
In order to maintain the overall charge neutrality of the system, $Z^{\rm c}c^{\rm c}=\vert Z^{\rm a}\vert c^{\rm a}$.
Moreover, counterions of the PE are the same species as those of the added salt.
The aforementioned salt, ion, and PE property relations apply throughout unless otherwise stated. 

\begin{figure}[t!]
\centering
  \includegraphics[width=1\linewidth]{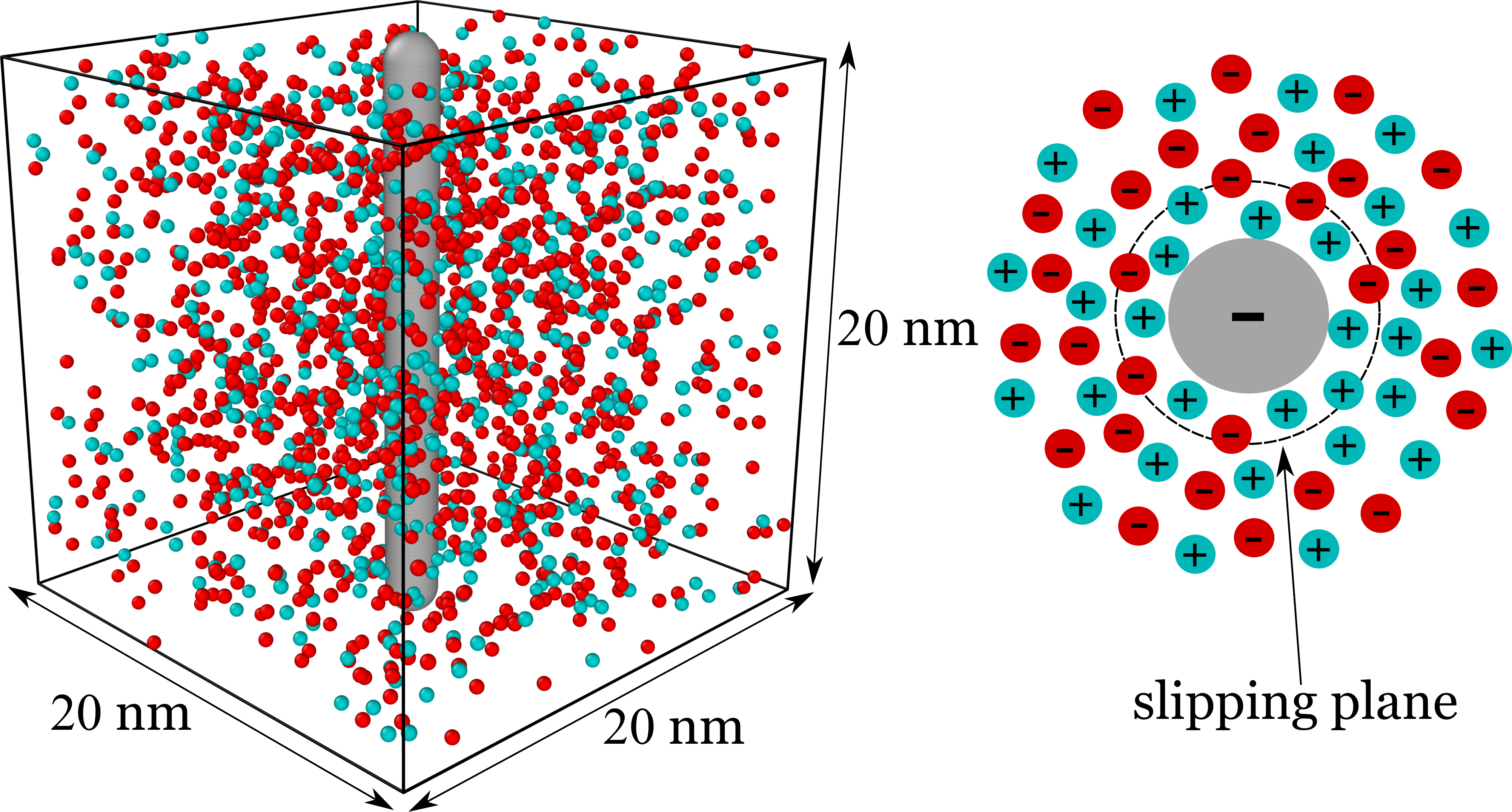}
 \caption{(a) Snapshot of the simulation box of size (20 nm)$^3$ with periodic boundary conditions containing one PE (gray), divalent counterions (cyan), and monovalent coions (red) at $Z^{\rm c}c^{\rm c}=0.25$ M. (b) Sketch showing a top view of ions around a negatively charged rod-like PE. The PE surface is surrounded by a layer composed of ions of opposite charge, beyond which a diffuse layer containing both negative and positive ions develops. The slipping plane is the interface between the mobile particles and the condensed ones. The $\zeta$ potential corresponds to the electrostatic potential at the slipping plane.}
 \label{box}
\end{figure}

Assuming angular homogeneity, we calculate the radial charge number density of the $i^{\rm th}$ ion species with a total number of ions $N^i$ as
\begin{equation}
    n^i(r)=\frac{\langle\sum_{k=1}^{N^i}Z^{i}\delta(\lvert\vect{r}_{\rm }-\vect{r}_{k}^{\,i}\lvert)\rangle_{t}}{2\pi r dr L_{z}},
    \label{eq3}
\end{equation}
where $\vect{r}=(x,y)$ is the distance vector on the $xy$ plane from the PE axis, $\langle\cdots\rangle_{t}$ the time average, $\vect{r}_{k}^{\,i}=(x_{k}^{\,i}, y_{k}^{\,i})$ are planar vectors pointing on single charges, $dr=0.01$ nm the cylindrical shell thickness, and $L_{z}$ the PE length. 

The cumulative radial charge density as a function of the radial distance from the PE axis, $r$ is expressed as
\begin{equation}
    \lambda_{\rm t}(r)=\lambda_0+{2\pi}e\sum_i\int_{0}^{r}n^{ i}(r^{\prime})r^{\prime}dr^{\prime}.
    \label{frac_eq}
\end{equation}
This represents the total charge inside a cylinder of radius $r$ normalized by the PE length. %
To obtain the radial electrostatic potential, $\phi(r)$, we employ Gauss' law
\begin{equation}
\phi(r)= \int_r^\infty\frac{\lambda_{\rm t}(r^\prime)}{2\pi r^\prime\varepsilon_0\varepsilon_{\rm r}}dr^\prime.
    \label{eq7}
\end{equation}
The mobility $\mu$ is obtained via the Helmholtz-Smoluchowski equation~\cite{von1903}
\begin{equation}
    \mu_{\rm }=\frac{\varepsilon_0\varepsilon_{\rm r} \zeta}{\eta},
\end{equation}
where $\eta=8.91\times 10^{-4}$ Pa s denotes the dynamic viscosity of water at room temperature, and the $\zeta$ potential is the electric potential at the radial distance corresponding to the slipping plane $r_{\rm sp}$ from the PE surface, i.e. $\zeta=\phi(\sigma^{\rm PE}/2+r_{\rm sp})$.
Although the exact value of $r_{\rm sp}$ is unknown, it can be approximated to be slightly larger than $\sigma$~\cite{joly2006, galla2014, raafatnia2014,yamaguchi2016}.
Here, we set $r_{\rm sp}=1.5\sigma$, corresponding to $0.75$ nm. A cylinder with a radius of $\sigma_{\rm PE}/2+1.5\sigma$ contains the majority of condensed counterions, as well as some coions. We have checked that our results are robust against small variations in $r_{\rm sp}$.

\begin{figure}[t!]
\centering
  \includegraphics[width=1\linewidth]{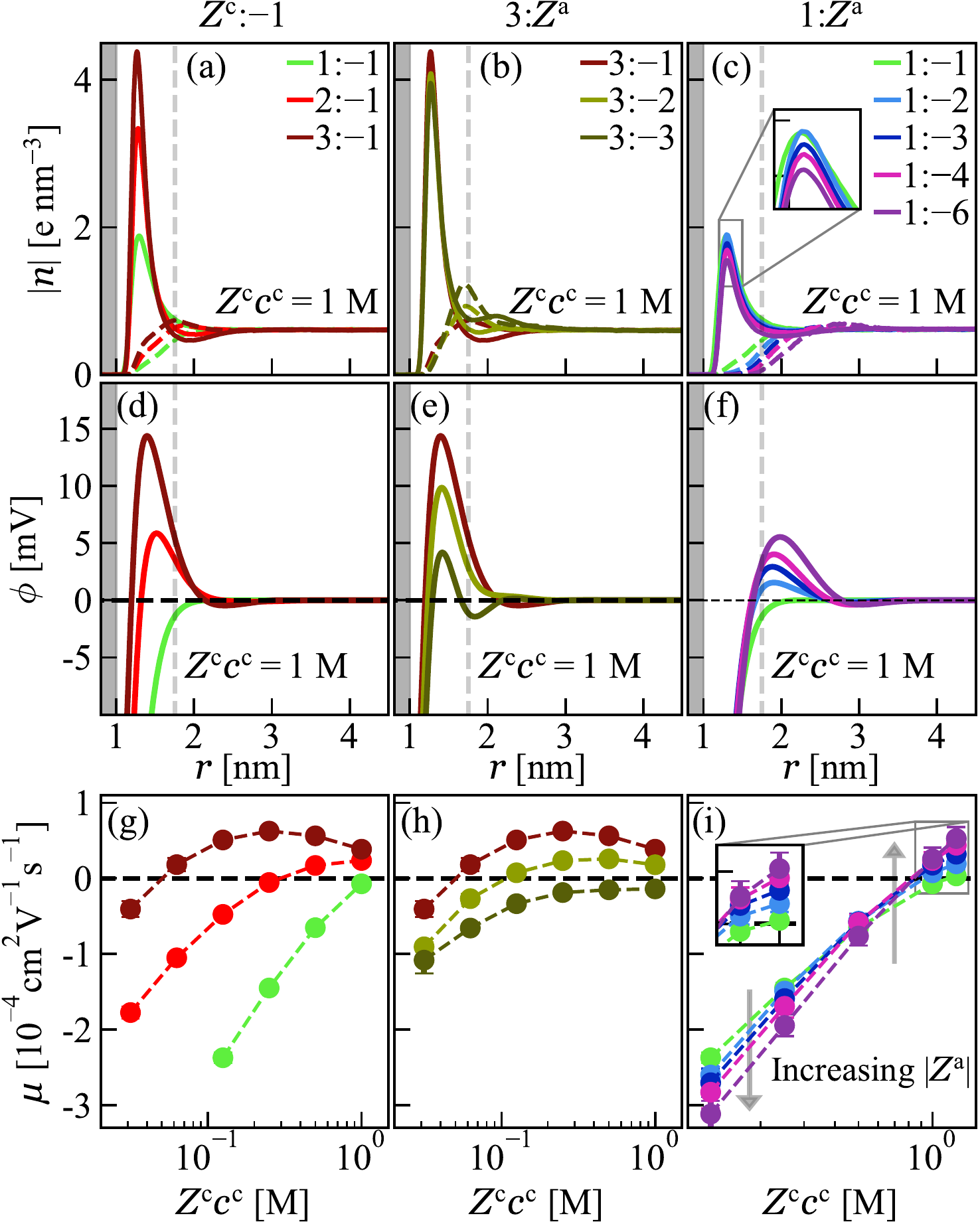}
 \caption{Counterion $\vert n^{\rm c}\vert$ (solid symbols) and coion $\vert n^{\rm a}\vert$ (open symbols) charge densities, defined by Eq.~(\ref{eq3}), as a function of the radial distance from the center of the PE at $Z^{\rm c}c^{\rm c}=1$ M in (a) $Z^{\rm c}$:$-1$, (b) 3:$Z^{\rm a}$, and (c) 1:$Z^{\rm a}$ salt solutions.
(d)-(f) The average electrostatic potential profiles (Eq.~(\ref{eq7})), for systems in (a)-(c).
 The gray bar indicates the PE with radius $1$ nm and dashed vertical lines the position of $r_{\rm sp}$.
 The EM as a function of counterion concentration $Z^{\rm c}c^{\rm c}$ in salt solutions with (g) $Z^{\rm c}$:$-1$, (h) 3:$Z^{\rm a}$, and (i) 1:$Z^{\rm a}$.
}
 \label{anion_valency}
\end{figure}

\begin{figure*}[htb!]
\centering
  \includegraphics[width=1\linewidth]{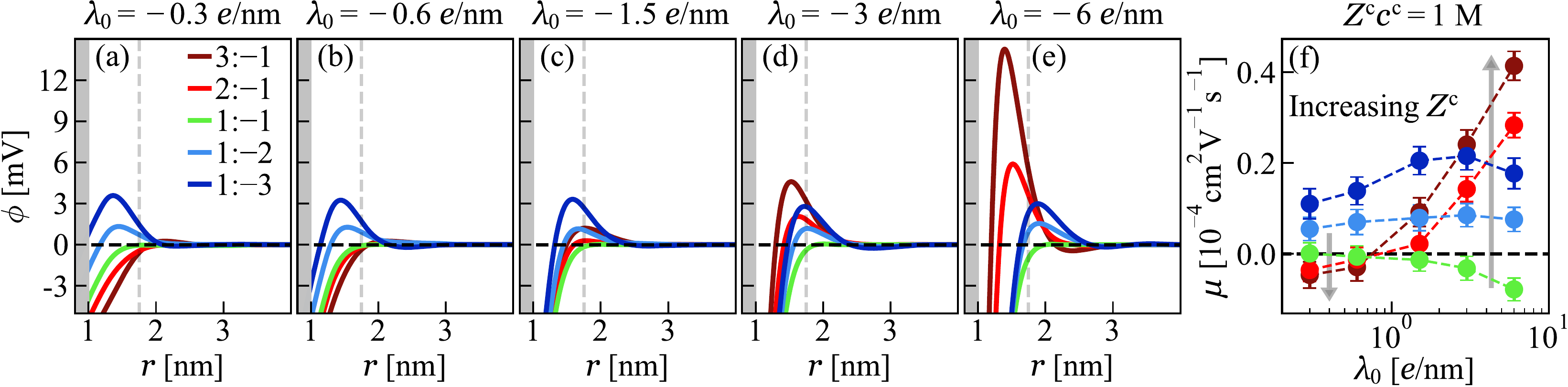}
 \caption{Panels (a)-(e) present the average electrostatic potential profiles (Eq.~(\ref{eq7})), as a function of the radial distance from the center of the PE at $Z^{\rm c}c^{\rm c}=1$ M, for different PE linear charge densities $\lambda_0$. 
 The gray bar indicates the PE with radius $1$ nm, and dashed vertical lines the position of $r_{\rm sp}$.
 (f) The EM as a function of $\lambda_0$ at $Z^{\rm c}c^{\rm c}=1$ M for different ion valencies.}
 \label{PE_charge}
\end{figure*}
\paragraph*{Results.}
\begin{figure}[b!]
\centering
  \includegraphics[width=1\linewidth]{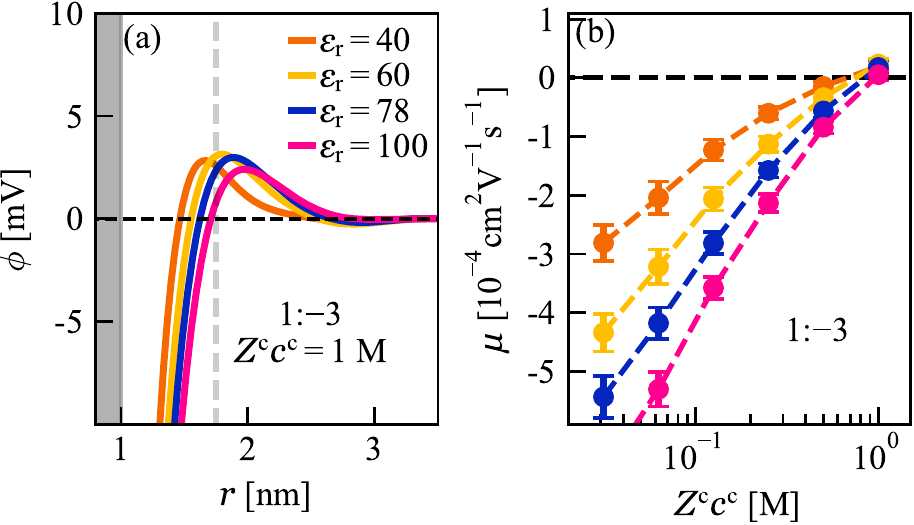}
 \caption{
(a) The average electrostatic potential profiles (Eq.~(\ref{eq7})), as a function of the radial distance from the center of the PE at $Z^{\rm c}c^{\rm c}=1$ M in $1$:$-3$ salt solutions for the varying relative dielectric constant of the solvent $\varepsilon_{\rm r}$.
 The gray bar indicates the PE with radius $1$ nm and dashed vertical lines the position of $r_{\rm sp}$.
 (b) The EM as a function of counterion concentration $Z^{\rm c}c^{\rm c}$ in $1$:$-3$ salt solutions for varying $\varepsilon_{\rm r}$.}
 \label{dielectric_medium}
\end{figure}

We first focus our attention on the effect of counterion valency.
Recently, in Ref.~\onlinecite{yang2023}, Yang {\it et al.} investigated the effect of counterion valency on DNA EM reversal. 
In their work, the DNA charge was neutralized by monovalent counterions, while the added salt contained multivalent cations and monovalent anions.
They demonstrated that the inversion of EM results from the overcompensation of DNA charge by multivalent cations, and the minimum critical amount of multivalent cations required for inversion decreases as cation valency increases.
As shown in Figs.~\ref{anion_valency}(a) and (g), increasing $Z^{\rm c}$ in $Z^{\rm c}$:$-1$ salts leads to enhanced EM reversal, which is due to the increased counterion condensation on the PE surface, consistent with the findings of Ref.~\onlinecite{yang2023}, as well as in the case of oppositely charged rods in charge-asymmetric salts~\cite{antila2017}.

In order to study the effect of $Z^{\rm a}$ on $\mu$, we first report charge density profiles for solutions of 3:$Z^{\rm a}$ salt, with $Z^{\rm a}=-1, -2$, and $-3$, shown in Fig.~\ref{anion_valency}(b).
As $\vert Z^{\rm a} \vert$ increases, we find that $n^{\rm c}$ does not change significantly, whereas a peak in $n^{\rm a}$ begins to appear.
This is because the strong Coulombic interaction between counterions and multivalent coions can condense coions into the counterion layer that initially triggered charge reversal~\cite{tanaka2003, wang2015, antila2017}.
The latter results in a considerable suppression of $\phi$, and consequently of $\mu$ (see Figs.~\ref{anion_valency}(e) and~\ref{anion_valency}(h)).
In this case, EM reversal is systematically weakened at all concentrations.
Somewhat surprisingly then, we find that in the symmetric $3$:$-3$ case, the EM reversal is completely suppressed despite the presence of multivalent counterions.
Conventionally, CR has been assumed to be strictly associated with high counterion valency~\cite{de1995, bloomfield1997, Shklovskii1999, Shklovskii1999-2, solis2000, nguyen2000, tanaka2004, besteman2007}.
Our findings, on the other hand, show that the valency of the coions also controls DNA-like PE mobility reversal, and can potentially prevent it from happening in symmetric multivalent solutions with $Z^{\rm c} = \vert Z^{\rm a}\vert=3$, consistent with Refs.~\citenum{tanaka2003, antila2017}.

We next consider the case in which the counterions are monovalent, but the coions are multivalent. The role of valency asymmetry in solutions of 1:$Z^{\rm a}$ is shown in Fig.~\ref{anion_valency}(i), where we observe that EM reversal occurs in 1:$Z^{\rm a}$ ($\vert Z^{\rm a} \vert\geq2$) salt solutions for relatively high concentrations, $c^{\rm c}\approx 1$ M. This surprising result can be attributed to the greater depletion of multivalent coions close to the PE surface as compared to their monovalent counterparts since the counterion density profiles do not change significantly (see Fig.~\ref{anion_valency}(c)).
These results are consistent with Monte Carlo simulations showing that CR occurs when a negatively charged flat surface is immersed in a highly concentrated solution of $1$:$-3$ salt~\cite{wang2015}.
On the other hand, when looking at intermediate-low concentrations ($c^{\rm c}\leq 0.5$ M), higher values of $\vert Z^{\rm a} \vert$ contribute to suppressing the EM reversal. Thus, there exists a concentration crossover at which the effect of coion valency switches its trend on the EM reversal.

In Fig.~\ref{PE_charge}, we show $\phi$ and $\mu$ for varying $\lambda_0$ and for different ion valencies at a fixed high salt concentration, namely $Z^{\rm c}c^{\rm c}=1$~M. As expected, no CR rises for symmetric monovalent ion salt systems, and for monovalent coions and multivalent counterions, EM reversal does not occur for the weakest line charge densities here. In the present case, the minimum value for reversal is $\lambda_0\approx -1.5 \ e/{\rm nm}$, and the peak in $\phi$ moves toward the PE surface as $\lambda_0$ increases. The most remarkable finding is that
for cases in which the counterions are monovalent but the coions are multivalent, there is a clear EM reversal for all the PE line charge densities considered here. 
Additionally, for increasing values of $\lambda_0$, the position of the peak in $\phi$ moves further away from the PE surface, indicating that a thicker layer of monovalent counterions is formed (see panels (a)-(e)).
In these cases, the height of this peak is almost insensitive to $\lambda_0$. 
Notably, as sketched by the arrows in panel (f), for small values of $\lambda_0$, higher values of $Z^{\rm c}$ decrease the value of $\mu$, whereas for higher values of $\lambda_0$, higher values of $Z^{\rm c}$ increase the value of $\mu$. This additional crossover underlines once more the non-monotonic response of the EM as a function of ion valencies.

Finally, we address the dependence of $\mu$ on the
solvent dielectric constant $\varepsilon_{\rm r}$. In panel (a) of Fig.~\ref{dielectric_medium} we show the electrostatic potential $\phi$.
We see that the peak position is monotonically shifted towards higher values of $r$, as $\varepsilon_{\rm r}$ increases. This can be explained by the fact that the electrostatic attraction of the counterions to the PE is weakened, and a greater concentration of counterions in the condensed layer is required to neutralize the PE charge.
The latter results in a monotonic increase of $\mu$ as a function of $\varepsilon_{\rm r}$, as shown in panel (b) of Fig.~\ref{dielectric_medium}. Interestingly, the EM is reversed for all considered values of $\varepsilon_{\rm r}$ at $Z^{\rm c}c^{\rm c}=1$ M, and its value does not change significantly since the PE charge is screened by the high salt concentration.
Nonetheless, the solvent screening effect is more effective at lower salt concentrations, and $\mu$ increases with increasing $\varepsilon_{\rm r}$. 

\paragraph*{Summary and Conclusions.}
In this work, we have highlighted the often overlooked effect of coion valency on the electrophoretic mobility of rod-like PEs for biologically relevant systems (model DNA). Contrary to the monotonic suppression effect found in previous works~\cite{tanaka2003, antila2017}, we find that such suppression only occurs at low and intermediate salt concentrations. In fact, in Fig.~\ref{anion_valency}(i), we observe a crossover (between approximately 0.5 M and 1 M), at which the effect of coion valency on the electrophoretic mobility is overturned. Specifically, at 1 M salt concentration, we observe electrophoretic mobility reversal for all but the $1$:$-1$ solution. Here, $\mu$ increases proportionally to $\vert Z^{\rm a} \vert$. The results for high concentrations are consistent with the findings on charge reversal for flat surfaces~\cite{wang2015}. Remarkably, electrophoretic mobility reversal at high concentrations of $1$:$Z^{\rm a}$ salt solutions, for $\vert Z^{\rm a}\vert\geq 2$, also happens for PEs with low linear charge densities where multivalent counterion-induced reversal is absent, as shown in Fig.~\ref{PE_charge}(f). However, we find a crossover in the effect of $Z^{\rm c}$ on $\mu$ for different values of $\lambda_0$. These non-monotonic effects provide an additional handle for controlling the interactions between charged molecules. Finally, we found that DNA electrophoretic mobility is reversed at 1 M concentration for $1$:$-3$ salts, independently from the value of the solvent dielectric constant.

\paragraph*{Acknowledgments:}
This work was supported by the Academy of Finland through its Centres of Excellence Programme (2022-2029, LIBER) under project no. 346111 (M.S.) and Academy of Finland project No. 353298 (T.A-N.). The work was also supported by Technology Industries of Finland Centennial Foundation TT2020 grant (T.A-N.).
We are grateful for the support by FinnCERES Materials Bioeconomy Ecosystem. Computational resources by CSC IT Centre for Finland and RAMI -- RawMatters Finland Infrastructure are also gratefully acknowledged.

\newpage
\bibliography{references}

\begin{thebibliography}{86}%
\makeatletter
\providecommand \@ifxundefined [1]{%
 \@ifx{#1\undefined}
}%
\providecommand \@ifnum [1]{%
 \ifnum #1\expandafter \@firstoftwo
 \else \expandafter \@secondoftwo
 \fi
}%
\providecommand \@ifx [1]{%
 \ifx #1\expandafter \@firstoftwo
 \else \expandafter \@secondoftwo
 \fi
}%
\providecommand \natexlab [1]{#1}%
\providecommand \enquote  [1]{``#1''}%
\providecommand \bibnamefont  [1]{#1}%
\providecommand \bibfnamefont [1]{#1}%
\providecommand \citenamefont [1]{#1}%
\providecommand \href@noop [0]{\@secondoftwo}%
\providecommand \href [0]{\begingroup \@sanitize@url \@href}%
\providecommand \@href[1]{\@@startlink{#1}\@@href}%
\providecommand \@@href[1]{\endgroup#1\@@endlink}%
\providecommand \@sanitize@url [0]{\catcode `\\12\catcode `\$12\catcode
  `\&12\catcode `\#12\catcode `\^12\catcode `\_12\catcode `\%12\relax}%
\providecommand \@@startlink[1]{}%
\providecommand \@@endlink[0]{}%
\providecommand \url  [0]{\begingroup\@sanitize@url \@url }%
\providecommand \@url [1]{\endgroup\@href {#1}{\urlprefix }}%
\providecommand \urlprefix  [0]{URL }%
\providecommand \Eprint [0]{\href }%
\providecommand \doibase [0]{http://dx.doi.org/}%
\providecommand \selectlanguage [0]{\@gobble}%
\providecommand \bibinfo  [0]{\@secondoftwo}%
\providecommand \bibfield  [0]{\@secondoftwo}%
\providecommand \translation [1]{[#1]}%
\providecommand \BibitemOpen [0]{}%
\providecommand \bibitemStop [0]{}%
\providecommand \bibitemNoStop [0]{.\EOS\space}%
\providecommand \EOS [0]{\spacefactor3000\relax}%
\providecommand \BibitemShut  [1]{\csname bibitem#1\endcsname}%
\let\auto@bib@innerbib\@empty
\bibitem [{\citenamefont {Strauss}\ \emph {et~al.}(1954)\citenamefont
  {Strauss}, \citenamefont {Gershfeld},\ and\ \citenamefont
  {Spiera}}]{strauss1954}%
  \BibitemOpen
  \bibfield  {author} {\bibinfo {author} {\bibfnamefont {U.~P.}\ \bibnamefont
  {Strauss}}, \bibinfo {author} {\bibfnamefont {N.~L.}\ \bibnamefont
  {Gershfeld}}, \ and\ \bibinfo {author} {\bibfnamefont {H.}~\bibnamefont
  {Spiera}},\ }\href@noop {} {\bibfield  {journal} {\bibinfo  {journal} {J. Am.
  Chem. Soc.}\ }\textbf {\bibinfo {volume} {76}},\ \bibinfo {pages} {5909}
  (\bibinfo {year} {1954})}\BibitemShut {NoStop}%
\bibitem [{\citenamefont {Olvera de~la Cruz}\ \emph {et~al.}(1995)\citenamefont
  {Olvera de~la Cruz}, \citenamefont {Belloni}, \citenamefont {Delsanti},
  \citenamefont {Dalbiez}, \citenamefont {Spalla},\ and\ \citenamefont
  {Drifford}}]{de1995}%
  \BibitemOpen
  \bibfield  {author} {\bibinfo {author} {\bibfnamefont {M.}~\bibnamefont
  {Olvera de~la Cruz}}, \bibinfo {author} {\bibfnamefont {L.}~\bibnamefont
  {Belloni}}, \bibinfo {author} {\bibfnamefont {M.}~\bibnamefont {Delsanti}},
  \bibinfo {author} {\bibfnamefont {J.}~\bibnamefont {Dalbiez}}, \bibinfo
  {author} {\bibfnamefont {O.}~\bibnamefont {Spalla}}, \ and\ \bibinfo {author}
  {\bibfnamefont {M.}~\bibnamefont {Drifford}},\ }\href@noop {} {\bibfield
  {journal} {\bibinfo  {journal} {J. Chem. Phys.}\ }\textbf {\bibinfo {volume}
  {103}},\ \bibinfo {pages} {5781} (\bibinfo {year} {1995})}\BibitemShut
  {NoStop}%
\bibitem [{\citenamefont {Grosberg}\ \emph {et~al.}(2002)\citenamefont
  {Grosberg}, \citenamefont {Nguyen},\ and\ \citenamefont
  {Shklovskii}}]{grosberg2002}%
  \BibitemOpen
  \bibfield  {author} {\bibinfo {author} {\bibfnamefont {A.~Y.}\ \bibnamefont
  {Grosberg}}, \bibinfo {author} {\bibfnamefont {T.}~\bibnamefont {Nguyen}}, \
  and\ \bibinfo {author} {\bibfnamefont {B.}~\bibnamefont {Shklovskii}},\
  }\href@noop {} {\bibfield  {journal} {\bibinfo  {journal} {Rev. Mod. Phys.}\
  }\textbf {\bibinfo {volume} {74}},\ \bibinfo {pages} {329} (\bibinfo {year}
  {2002})}\BibitemShut {NoStop}%
\bibitem [{\citenamefont {Decher}(1997)}]{decher1997}%
  \BibitemOpen
  \bibfield  {author} {\bibinfo {author} {\bibfnamefont {G.}~\bibnamefont
  {Decher}},\ }\href@noop {} {\bibfield  {journal} {\bibinfo  {journal}
  {science}\ }\textbf {\bibinfo {volume} {277}},\ \bibinfo {pages} {1232}
  (\bibinfo {year} {1997})}\BibitemShut {NoStop}%
\bibitem [{\citenamefont {Koltover}\ \emph {et~al.}(1998)\citenamefont
  {Koltover}, \citenamefont {Salditt}, \citenamefont {Rädler},\ and\
  \citenamefont {Safinya}}]{koltover1998}%
  \BibitemOpen
  \bibfield  {author} {\bibinfo {author} {\bibfnamefont {I.}~\bibnamefont
  {Koltover}}, \bibinfo {author} {\bibfnamefont {T.}~\bibnamefont {Salditt}},
  \bibinfo {author} {\bibfnamefont {J.~O.}\ \bibnamefont {Rädler}}, \ and\
  \bibinfo {author} {\bibfnamefont {C.~R.}\ \bibnamefont {Safinya}},\ }\href
  {\doibase 10.1126/science.281.5373.78} {\bibfield  {journal} {\bibinfo
  {journal} {Science}\ }\textbf {\bibinfo {volume} {281}},\ \bibinfo {pages}
  {78} (\bibinfo {year} {1998})}\BibitemShut {NoStop}%
\bibitem [{\citenamefont {Besteman}\ \emph
  {et~al.}(2007{\natexlab{a}})\citenamefont {Besteman}, \citenamefont
  {Van~Eijk},\ and\ \citenamefont {Lemay}}]{besteman2007}%
  \BibitemOpen
  \bibfield  {author} {\bibinfo {author} {\bibfnamefont {K.}~\bibnamefont
  {Besteman}}, \bibinfo {author} {\bibfnamefont {K.}~\bibnamefont {Van~Eijk}},
  \ and\ \bibinfo {author} {\bibfnamefont {S.}~\bibnamefont {Lemay}},\
  }\href@noop {} {\bibfield  {journal} {\bibinfo  {journal} {Nat. Phys.}\
  }\textbf {\bibinfo {volume} {3}},\ \bibinfo {pages} {641} (\bibinfo {year}
  {2007}{\natexlab{a}})}\BibitemShut {NoStop}%
\bibitem [{\citenamefont {Besteman}\ \emph
  {et~al.}(2007{\natexlab{b}})\citenamefont {Besteman}, \citenamefont {Hage},
  \citenamefont {Dekker},\ and\ \citenamefont {Lemay}}]{besteman2007-2}%
  \BibitemOpen
  \bibfield  {author} {\bibinfo {author} {\bibfnamefont {K.}~\bibnamefont
  {Besteman}}, \bibinfo {author} {\bibfnamefont {S.}~\bibnamefont {Hage}},
  \bibinfo {author} {\bibfnamefont {N.}~\bibnamefont {Dekker}}, \ and\ \bibinfo
  {author} {\bibfnamefont {S.}~\bibnamefont {Lemay}},\ }\href@noop {}
  {\bibfield  {journal} {\bibinfo  {journal} {Phys. Rev. Lett.}\ }\textbf
  {\bibinfo {volume} {98}},\ \bibinfo {pages} {058103} (\bibinfo {year}
  {2007}{\natexlab{b}})}\BibitemShut {NoStop}%
\bibitem [{\citenamefont {Sanchez-Ruiz}\ and\ \citenamefont
  {Makhatadze}(2001)}]{sanchez2001}%
  \BibitemOpen
  \bibfield  {author} {\bibinfo {author} {\bibfnamefont {J.~M.}\ \bibnamefont
  {Sanchez-Ruiz}}\ and\ \bibinfo {author} {\bibfnamefont {G.~I.}\ \bibnamefont
  {Makhatadze}},\ }\href@noop {} {\bibfield  {journal} {\bibinfo  {journal}
  {Trends Biotechnol.}\ }\textbf {\bibinfo {volume} {19}},\ \bibinfo {pages}
  {132} (\bibinfo {year} {2001})}\BibitemShut {NoStop}%
\bibitem [{\citenamefont {Hotze}\ \emph {et~al.}(2010)\citenamefont {Hotze},
  \citenamefont {Phenrat},\ and\ \citenamefont {Lowry}}]{hotze2010}%
  \BibitemOpen
  \bibfield  {author} {\bibinfo {author} {\bibfnamefont {E.~M.}\ \bibnamefont
  {Hotze}}, \bibinfo {author} {\bibfnamefont {T.}~\bibnamefont {Phenrat}}, \
  and\ \bibinfo {author} {\bibfnamefont {G.~V.}\ \bibnamefont {Lowry}},\
  }\href@noop {} {\bibfield  {journal} {\bibinfo  {journal} {J. Environ.
  Qual.}\ }\textbf {\bibinfo {volume} {39}},\ \bibinfo {pages} {1909} (\bibinfo
  {year} {2010})}\BibitemShut {NoStop}%
\bibitem [{\citenamefont {Besteman}\ \emph {et~al.}(2004)\citenamefont
  {Besteman}, \citenamefont {Zevenbergen}, \citenamefont {Heering},\ and\
  \citenamefont {Lemay}}]{besteman2004}%
  \BibitemOpen
  \bibfield  {author} {\bibinfo {author} {\bibfnamefont {K.}~\bibnamefont
  {Besteman}}, \bibinfo {author} {\bibfnamefont {M.~A.}\ \bibnamefont
  {Zevenbergen}}, \bibinfo {author} {\bibfnamefont {H.~A.}\ \bibnamefont
  {Heering}}, \ and\ \bibinfo {author} {\bibfnamefont {S.~G.}\ \bibnamefont
  {Lemay}},\ }\href@noop {} {\bibfield  {journal} {\bibinfo  {journal} {Phys.
  Rev. Lett.}\ }\textbf {\bibinfo {volume} {93}},\ \bibinfo {pages} {170802}
  (\bibinfo {year} {2004})}\BibitemShut {NoStop}%
\bibitem [{\citenamefont {Besteman}\ \emph {et~al.}(2005)\citenamefont
  {Besteman}, \citenamefont {Zevenbergen},\ and\ \citenamefont
  {Lemay}}]{besteman2005}%
  \BibitemOpen
  \bibfield  {author} {\bibinfo {author} {\bibfnamefont {K.}~\bibnamefont
  {Besteman}}, \bibinfo {author} {\bibfnamefont {M.}~\bibnamefont
  {Zevenbergen}}, \ and\ \bibinfo {author} {\bibfnamefont {S.}~\bibnamefont
  {Lemay}},\ }\href@noop {} {\bibfield  {journal} {\bibinfo  {journal} {Phys.
  Rev. E}\ }\textbf {\bibinfo {volume} {72}},\ \bibinfo {pages} {061501}
  (\bibinfo {year} {2005})}\BibitemShut {NoStop}%
\bibitem [{\citenamefont {Zomer}\ \emph {et~al.}(2016)\citenamefont {Zomer},
  \citenamefont {Neufeldt}, \citenamefont {Xu}, \citenamefont {Ahrends},
  \citenamefont {Bossio}, \citenamefont {Trabucco}, \citenamefont
  {Van~Noordwijk},\ and\ \citenamefont {Wang}}]{zomer2016}%
  \BibitemOpen
  \bibfield  {author} {\bibinfo {author} {\bibfnamefont {R.~J.}\ \bibnamefont
  {Zomer}}, \bibinfo {author} {\bibfnamefont {H.}~\bibnamefont {Neufeldt}},
  \bibinfo {author} {\bibfnamefont {J.}~\bibnamefont {Xu}}, \bibinfo {author}
  {\bibfnamefont {A.}~\bibnamefont {Ahrends}}, \bibinfo {author} {\bibfnamefont
  {D.}~\bibnamefont {Bossio}}, \bibinfo {author} {\bibfnamefont
  {A.}~\bibnamefont {Trabucco}}, \bibinfo {author} {\bibfnamefont
  {M.}~\bibnamefont {Van~Noordwijk}}, \ and\ \bibinfo {author} {\bibfnamefont
  {M.}~\bibnamefont {Wang}},\ }\href@noop {} {\bibfield  {journal} {\bibinfo
  {journal} {Sci. Rep.}\ }\textbf {\bibinfo {volume} {6}},\ \bibinfo {pages}
  {29987} (\bibinfo {year} {2016})}\BibitemShut {NoStop}%
\bibitem [{\citenamefont {Wang}\ \emph {et~al.}(2018)\citenamefont {Wang},
  \citenamefont {Wang}, \citenamefont {Gao},\ and\ \citenamefont
  {Yang}}]{wang2018}%
  \BibitemOpen
  \bibfield  {author} {\bibinfo {author} {\bibfnamefont {Y.}~\bibnamefont
  {Wang}}, \bibinfo {author} {\bibfnamefont {R.}~\bibnamefont {Wang}}, \bibinfo
  {author} {\bibfnamefont {T.}~\bibnamefont {Gao}}, \ and\ \bibinfo {author}
  {\bibfnamefont {G.}~\bibnamefont {Yang}},\ }\href@noop {} {\bibfield
  {journal} {\bibinfo  {journal} {Polymers}\ }\textbf {\bibinfo {volume}
  {10}},\ \bibinfo {pages} {244} (\bibinfo {year} {2018})}\BibitemShut
  {NoStop}%
\bibitem [{\citenamefont {Kandu{\v{c}}}\ \emph {et~al.}(2010)\citenamefont
  {Kandu{\v{c}}}, \citenamefont {Naji}, \citenamefont {Forsman},\ and\
  \citenamefont {Podgornik}}]{kanduvc2010}%
  \BibitemOpen
  \bibfield  {author} {\bibinfo {author} {\bibfnamefont {M.}~\bibnamefont
  {Kandu{\v{c}}}}, \bibinfo {author} {\bibfnamefont {A.}~\bibnamefont {Naji}},
  \bibinfo {author} {\bibfnamefont {J.}~\bibnamefont {Forsman}}, \ and\
  \bibinfo {author} {\bibfnamefont {R.}~\bibnamefont {Podgornik}},\ }\href@noop
  {} {\bibfield  {journal} {\bibinfo  {journal} {J. Chem. Phys.}\ }\textbf
  {\bibinfo {volume} {132}},\ \bibinfo {pages} {124701} (\bibinfo {year}
  {2010})}\BibitemShut {NoStop}%
\bibitem [{\citenamefont {Kandu{\v{c}}}\ \emph {et~al.}(2011)\citenamefont
  {Kandu{\v{c}}}, \citenamefont {Naji}, \citenamefont {Forsman},\ and\
  \citenamefont {Podgornik}}]{kanduvc2011}%
  \BibitemOpen
  \bibfield  {author} {\bibinfo {author} {\bibfnamefont {M.}~\bibnamefont
  {Kandu{\v{c}}}}, \bibinfo {author} {\bibfnamefont {A.}~\bibnamefont {Naji}},
  \bibinfo {author} {\bibfnamefont {J.}~\bibnamefont {Forsman}}, \ and\
  \bibinfo {author} {\bibfnamefont {R.}~\bibnamefont {Podgornik}},\ }\href@noop
  {} {\bibfield  {journal} {\bibinfo  {journal} {Phys. Rev. E}\ }\textbf
  {\bibinfo {volume} {84}},\ \bibinfo {pages} {011502} (\bibinfo {year}
  {2011})}\BibitemShut {NoStop}%
\bibitem [{\citenamefont {Buyukdagli}(2020)}]{buyukdagli2020}%
  \BibitemOpen
  \bibfield  {author} {\bibinfo {author} {\bibfnamefont {S.}~\bibnamefont
  {Buyukdagli}},\ }\href@noop {} {\bibfield  {journal} {\bibinfo  {journal} {J.
  Phys. Chem. B}\ }\textbf {\bibinfo {volume} {124}},\ \bibinfo {pages} {11299}
  (\bibinfo {year} {2020})}\BibitemShut {NoStop}%
\bibitem [{\citenamefont {Buyukdagli}(2021)}]{buyukdagli2021}%
  \BibitemOpen
  \bibfield  {author} {\bibinfo {author} {\bibfnamefont {S.}~\bibnamefont
  {Buyukdagli}},\ }\href@noop {} {\bibfield  {journal} {\bibinfo  {journal}
  {Langmuir}\ }\textbf {\bibinfo {volume} {38}},\ \bibinfo {pages} {122}
  (\bibinfo {year} {2021})}\BibitemShut {NoStop}%
\bibitem [{\citenamefont {Yang}\ \emph {et~al.}(2023)\citenamefont {Yang},
  \citenamefont {Buyukdagli}, \citenamefont {Scacchi}, \citenamefont
  {Sammalkorpi},\ and\ \citenamefont {Ala-Nissila}}]{yang2023}%
  \BibitemOpen
  \bibfield  {author} {\bibinfo {author} {\bibfnamefont {X.}~\bibnamefont
  {Yang}}, \bibinfo {author} {\bibfnamefont {S.}~\bibnamefont {Buyukdagli}},
  \bibinfo {author} {\bibfnamefont {A.}~\bibnamefont {Scacchi}}, \bibinfo
  {author} {\bibfnamefont {M.}~\bibnamefont {Sammalkorpi}}, \ and\ \bibinfo
  {author} {\bibfnamefont {T.}~\bibnamefont {Ala-Nissila}},\ }\href {\doibase
  10.1103/PhysRevE.107.034503} {\bibfield  {journal} {\bibinfo  {journal}
  {Phys. Rev. E}\ }\textbf {\bibinfo {volume} {107}},\ \bibinfo {pages}
  {034503} (\bibinfo {year} {2023})}\BibitemShut {NoStop}%
\bibitem [{\citenamefont {Deserno}\ \emph {et~al.}(2001)\citenamefont
  {Deserno}, \citenamefont {Jim{\'e}nez-{\'A}ngeles}, \citenamefont {Holm},\
  and\ \citenamefont {Lozada-Cassou}}]{deserno2001}%
  \BibitemOpen
  \bibfield  {author} {\bibinfo {author} {\bibfnamefont {M.}~\bibnamefont
  {Deserno}}, \bibinfo {author} {\bibfnamefont {F.}~\bibnamefont
  {Jim{\'e}nez-{\'A}ngeles}}, \bibinfo {author} {\bibfnamefont
  {C.}~\bibnamefont {Holm}}, \ and\ \bibinfo {author} {\bibfnamefont
  {M.}~\bibnamefont {Lozada-Cassou}},\ }\href@noop {} {\bibfield  {journal}
  {\bibinfo  {journal} {J. Phys. Chem. B}\ }\textbf {\bibinfo {volume} {105}},\
  \bibinfo {pages} {10983} (\bibinfo {year} {2001})}\BibitemShut {NoStop}%
\bibitem [{\citenamefont {Luan}\ and\ \citenamefont
  {Aksimentiev}(2010)}]{luan2010}%
  \BibitemOpen
  \bibfield  {author} {\bibinfo {author} {\bibfnamefont {B.}~\bibnamefont
  {Luan}}\ and\ \bibinfo {author} {\bibfnamefont {A.}~\bibnamefont
  {Aksimentiev}},\ }\href@noop {} {\bibfield  {journal} {\bibinfo  {journal}
  {Soft Matter}\ }\textbf {\bibinfo {volume} {6}},\ \bibinfo {pages} {243}
  (\bibinfo {year} {2010})}\BibitemShut {NoStop}%
\bibitem [{\citenamefont {Tanaka}(2003)}]{tanaka2003}%
  \BibitemOpen
  \bibfield  {author} {\bibinfo {author} {\bibfnamefont {M.}~\bibnamefont
  {Tanaka}},\ }\href@noop {} {\bibfield  {journal} {\bibinfo  {journal} {Phys.
  Rev. E}\ }\textbf {\bibinfo {volume} {68}},\ \bibinfo {pages} {061501}
  (\bibinfo {year} {2003})}\BibitemShut {NoStop}%
\bibitem [{\citenamefont {Nguyen}\ \emph {et~al.}(2000)\citenamefont {Nguyen},
  \citenamefont {Rouzina},\ and\ \citenamefont {Shklovskii}}]{nguyen2000}%
  \BibitemOpen
  \bibfield  {author} {\bibinfo {author} {\bibfnamefont {T.~T.}\ \bibnamefont
  {Nguyen}}, \bibinfo {author} {\bibfnamefont {I.}~\bibnamefont {Rouzina}}, \
  and\ \bibinfo {author} {\bibfnamefont {B.~I.}\ \bibnamefont {Shklovskii}},\
  }\href@noop {} {\bibfield  {journal} {\bibinfo  {journal} {J. Chem. Phys.}\
  }\textbf {\bibinfo {volume} {112}},\ \bibinfo {pages} {2562} (\bibinfo {year}
  {2000})}\BibitemShut {NoStop}%
\bibitem [{\citenamefont {Becker}\ \emph {et~al.}(2012)\citenamefont {Becker},
  \citenamefont {Henzler}, \citenamefont {Welsch}, \citenamefont {Ballauff},\
  and\ \citenamefont {Borisov}}]{becker2012}%
  \BibitemOpen
  \bibfield  {author} {\bibinfo {author} {\bibfnamefont {A.~L.}\ \bibnamefont
  {Becker}}, \bibinfo {author} {\bibfnamefont {K.}~\bibnamefont {Henzler}},
  \bibinfo {author} {\bibfnamefont {N.}~\bibnamefont {Welsch}}, \bibinfo
  {author} {\bibfnamefont {M.}~\bibnamefont {Ballauff}}, \ and\ \bibinfo
  {author} {\bibfnamefont {O.}~\bibnamefont {Borisov}},\ }\href@noop {}
  {\bibfield  {journal} {\bibinfo  {journal} {Curr. Opin. Colloid Interface
  Sci.}\ }\textbf {\bibinfo {volume} {17}},\ \bibinfo {pages} {90} (\bibinfo
  {year} {2012})}\BibitemShut {NoStop}%
\bibitem [{\citenamefont {Szilagyi}\ \emph {et~al.}(2014)\citenamefont
  {Szilagyi}, \citenamefont {Trefalt}, \citenamefont {Tiraferri}, \citenamefont
  {Maroni},\ and\ \citenamefont {Borkovec}}]{szilagyi2014}%
  \BibitemOpen
  \bibfield  {author} {\bibinfo {author} {\bibfnamefont {I.}~\bibnamefont
  {Szilagyi}}, \bibinfo {author} {\bibfnamefont {G.}~\bibnamefont {Trefalt}},
  \bibinfo {author} {\bibfnamefont {A.}~\bibnamefont {Tiraferri}}, \bibinfo
  {author} {\bibfnamefont {P.}~\bibnamefont {Maroni}}, \ and\ \bibinfo {author}
  {\bibfnamefont {M.}~\bibnamefont {Borkovec}},\ }\href@noop {} {\bibfield
  {journal} {\bibinfo  {journal} {Soft Matter}\ }\textbf {\bibinfo {volume}
  {10}},\ \bibinfo {pages} {2479} (\bibinfo {year} {2014})}\BibitemShut
  {NoStop}%
\bibitem [{\citenamefont {Antila}\ and\ \citenamefont
  {Sammalkorpi}(2014)}]{antila2014}%
  \BibitemOpen
  \bibfield  {author} {\bibinfo {author} {\bibfnamefont {H.~S.}\ \bibnamefont
  {Antila}}\ and\ \bibinfo {author} {\bibfnamefont {M.}~\bibnamefont
  {Sammalkorpi}},\ }\href@noop {} {\bibfield  {journal} {\bibinfo  {journal}
  {J. Phys. Chem. B}\ }\textbf {\bibinfo {volume} {118}},\ \bibinfo {pages}
  {3226} (\bibinfo {year} {2014})}\BibitemShut {NoStop}%
\bibitem [{\citenamefont {Antila}\ \emph {et~al.}(2015)\citenamefont {Antila},
  \citenamefont {H\"ark\"onen},\ and\ \citenamefont
  {Sammalkorpi}}]{antila2015b}%
  \BibitemOpen
  \bibfield  {author} {\bibinfo {author} {\bibfnamefont {H.~S.}\ \bibnamefont
  {Antila}}, \bibinfo {author} {\bibfnamefont {M.}~\bibnamefont
  {H\"ark\"onen}}, \ and\ \bibinfo {author} {\bibfnamefont {M.}~\bibnamefont
  {Sammalkorpi}},\ }\href {\doibase 10.1039/C4CP04967E} {\bibfield  {journal}
  {\bibinfo  {journal} {Phys. Chem. Chem. Phys.}\ }\textbf {\bibinfo {volume}
  {17}},\ \bibinfo {pages} {5279} (\bibinfo {year} {2015})}\BibitemShut
  {NoStop}%
\bibitem [{\citenamefont {Antila}\ \emph {et~al.}(2016)\citenamefont {Antila},
  \citenamefont {Van~Tassel},\ and\ \citenamefont {Sammalkorpi}}]{Antila2016}%
  \BibitemOpen
  \bibfield  {author} {\bibinfo {author} {\bibfnamefont {H.~S.}\ \bibnamefont
  {Antila}}, \bibinfo {author} {\bibfnamefont {P.~R.}\ \bibnamefont
  {Van~Tassel}}, \ and\ \bibinfo {author} {\bibfnamefont {M.}~\bibnamefont
  {Sammalkorpi}},\ }\href@noop {} {\bibfield  {journal} {\bibinfo  {journal}
  {Phys. Rev. E}\ }\textbf {\bibinfo {volume} {93}},\ \bibinfo {pages} {022602}
  (\bibinfo {year} {2016})}\BibitemShut {NoStop}%
\bibitem [{\citenamefont {Antila}\ \emph {et~al.}(2017)\citenamefont {Antila},
  \citenamefont {Van~Tassel},\ and\ \citenamefont {Sammalkorpi}}]{antila2017}%
  \BibitemOpen
  \bibfield  {author} {\bibinfo {author} {\bibfnamefont {H.~S.}\ \bibnamefont
  {Antila}}, \bibinfo {author} {\bibfnamefont {P.~R.}\ \bibnamefont
  {Van~Tassel}}, \ and\ \bibinfo {author} {\bibfnamefont {M.}~\bibnamefont
  {Sammalkorpi}},\ }\href@noop {} {\bibfield  {journal} {\bibinfo  {journal}
  {J. Chem. Phys.}\ }\textbf {\bibinfo {volume} {147}},\ \bibinfo {pages}
  {124901} (\bibinfo {year} {2017})}\BibitemShut {NoStop}%
\bibitem [{\citenamefont {Vahid}\ \emph {et~al.}(2022)\citenamefont {Vahid},
  \citenamefont {Scacchi}, \citenamefont {Yang}, \citenamefont {Ala-Nissila},\
  and\ \citenamefont {Sammalkorpi}}]{vahid2022}%
  \BibitemOpen
  \bibfield  {author} {\bibinfo {author} {\bibfnamefont {H.}~\bibnamefont
  {Vahid}}, \bibinfo {author} {\bibfnamefont {A.}~\bibnamefont {Scacchi}},
  \bibinfo {author} {\bibfnamefont {X.}~\bibnamefont {Yang}}, \bibinfo {author}
  {\bibfnamefont {T.}~\bibnamefont {Ala-Nissila}}, \ and\ \bibinfo {author}
  {\bibfnamefont {M.}~\bibnamefont {Sammalkorpi}},\ }\href@noop {} {\bibfield
  {journal} {\bibinfo  {journal} {J. Chem. Phys.}\ }\textbf {\bibinfo {volume}
  {156}},\ \bibinfo {pages} {214906} (\bibinfo {year} {2022})}\BibitemShut
  {NoStop}%
\bibitem [{\citenamefont {Vahid}\ \emph {et~al.}(2023)\citenamefont {Vahid},
  \citenamefont {Scacchi}, \citenamefont {Sammalkorpi},\ and\ \citenamefont
  {Ala-Nissila}}]{vahid2023}%
  \BibitemOpen
  \bibfield  {author} {\bibinfo {author} {\bibfnamefont {H.}~\bibnamefont
  {Vahid}}, \bibinfo {author} {\bibfnamefont {A.}~\bibnamefont {Scacchi}},
  \bibinfo {author} {\bibfnamefont {M.}~\bibnamefont {Sammalkorpi}}, \ and\
  \bibinfo {author} {\bibfnamefont {T.}~\bibnamefont {Ala-Nissila}},\ }\href
  {\doibase 10.1103/PhysRevLett.130.158202} {\bibfield  {journal} {\bibinfo
  {journal} {Phys. Rev. Lett.}\ }\textbf {\bibinfo {volume} {130}},\ \bibinfo
  {pages} {158202} (\bibinfo {year} {2023})}\BibitemShut {NoStop}%
\bibitem [{\citenamefont {Yang}\ \emph {et~al.}(2022)\citenamefont {Yang},
  \citenamefont {Scacchi}, \citenamefont {Vahid}, \citenamefont {Sammalkorpi},\
  and\ \citenamefont {Ala-Nissila}}]{yang2022}%
  \BibitemOpen
  \bibfield  {author} {\bibinfo {author} {\bibfnamefont {X.}~\bibnamefont
  {Yang}}, \bibinfo {author} {\bibfnamefont {A.}~\bibnamefont {Scacchi}},
  \bibinfo {author} {\bibfnamefont {H.}~\bibnamefont {Vahid}}, \bibinfo
  {author} {\bibfnamefont {M.}~\bibnamefont {Sammalkorpi}}, \ and\ \bibinfo
  {author} {\bibfnamefont {T.}~\bibnamefont {Ala-Nissila}},\ }\href@noop {}
  {\bibfield  {journal} {\bibinfo  {journal} {Phys. Chem. Chem. Phys.}\
  }\textbf {\bibinfo {volume} {24}},\ \bibinfo {pages} {21112} (\bibinfo {year}
  {2022})}\BibitemShut {NoStop}%
\bibitem [{\citenamefont {Bloomfield}(1997)}]{bloomfield1997}%
  \BibitemOpen
  \bibfield  {author} {\bibinfo {author} {\bibfnamefont {V.~A.}\ \bibnamefont
  {Bloomfield}},\ }\href@noop {} {\bibfield  {journal} {\bibinfo  {journal}
  {Biopolymers}\ }\textbf {\bibinfo {volume} {44}},\ \bibinfo {pages} {269}
  (\bibinfo {year} {1997})}\BibitemShut {NoStop}%
\bibitem [{\citenamefont {Shklovskii}(1999{\natexlab{a}})}]{Shklovskii1999}%
  \BibitemOpen
  \bibfield  {author} {\bibinfo {author} {\bibfnamefont {B.~I.}\ \bibnamefont
  {Shklovskii}},\ }\href@noop {} {\bibfield  {journal} {\bibinfo  {journal}
  {Phys. Rev. E}\ }\textbf {\bibinfo {volume} {60}},\ \bibinfo {pages} {5802}
  (\bibinfo {year} {1999}{\natexlab{a}})}\BibitemShut {NoStop}%
\bibitem [{\citenamefont {Shklovskii}(1999{\natexlab{b}})}]{Shklovskii1999-2}%
  \BibitemOpen
  \bibfield  {author} {\bibinfo {author} {\bibfnamefont {B.~I.}\ \bibnamefont
  {Shklovskii}},\ }\href {\doibase 10.1103/PhysRevLett.82.3268} {\bibfield
  {journal} {\bibinfo  {journal} {Phys. Rev. Lett.}\ }\textbf {\bibinfo
  {volume} {82}},\ \bibinfo {pages} {3268} (\bibinfo {year}
  {1999}{\natexlab{b}})}\BibitemShut {NoStop}%
\bibitem [{\citenamefont {Solis}\ and\ \citenamefont {Olvera de~la
  Cruz}(2000)}]{solis2000}%
  \BibitemOpen
  \bibfield  {author} {\bibinfo {author} {\bibfnamefont {F.~J.}\ \bibnamefont
  {Solis}}\ and\ \bibinfo {author} {\bibfnamefont {M.}~\bibnamefont {Olvera
  de~la Cruz}},\ }\href@noop {} {\bibfield  {journal} {\bibinfo  {journal} {J.
  Chem. Phys.}\ }\textbf {\bibinfo {volume} {112}},\ \bibinfo {pages} {2030}
  (\bibinfo {year} {2000})}\BibitemShut {NoStop}%
\bibitem [{\citenamefont {Tanaka}(2004)}]{tanaka2004}%
  \BibitemOpen
  \bibfield  {author} {\bibinfo {author} {\bibfnamefont {M.}~\bibnamefont
  {Tanaka}},\ }\href@noop {} {\bibfield  {journal} {\bibinfo  {journal} {J.
  Phys. Condens. Matter}\ }\textbf {\bibinfo {volume} {16}},\ \bibinfo {pages}
  {S2127} (\bibinfo {year} {2004})}\BibitemShut {NoStop}%
\bibitem [{\citenamefont {Gonzales-Tovar}\ \emph {et~al.}(1985)\citenamefont
  {Gonzales-Tovar}, \citenamefont {Lozada-Cassou},\ and\ \citenamefont
  {Henderson}}]{gonzales1985}%
  \BibitemOpen
  \bibfield  {author} {\bibinfo {author} {\bibfnamefont {E.}~\bibnamefont
  {Gonzales-Tovar}}, \bibinfo {author} {\bibfnamefont {M.}~\bibnamefont
  {Lozada-Cassou}}, \ and\ \bibinfo {author} {\bibfnamefont {D.}~\bibnamefont
  {Henderson}},\ }\href@noop {} {\bibfield  {journal} {\bibinfo  {journal} {J.
  Chem. Phys.}\ }\textbf {\bibinfo {volume} {83}},\ \bibinfo {pages} {361}
  (\bibinfo {year} {1985})}\BibitemShut {NoStop}%
\bibitem [{\citenamefont {Jim{\'e}nez-{\'A}ngeles}\ \emph
  {et~al.}(2006)\citenamefont {Jim{\'e}nez-{\'A}ngeles}, \citenamefont
  {Odriozola},\ and\ \citenamefont {Lozada-Cassou}}]{jimenez2006}%
  \BibitemOpen
  \bibfield  {author} {\bibinfo {author} {\bibfnamefont {F.}~\bibnamefont
  {Jim{\'e}nez-{\'A}ngeles}}, \bibinfo {author} {\bibfnamefont
  {G.}~\bibnamefont {Odriozola}}, \ and\ \bibinfo {author} {\bibfnamefont
  {M.}~\bibnamefont {Lozada-Cassou}},\ }\href@noop {} {\bibfield  {journal}
  {\bibinfo  {journal} {J. Chem. Phys.}\ }\textbf {\bibinfo {volume} {124}},\
  \bibinfo {pages} {134902} (\bibinfo {year} {2006})}\BibitemShut {NoStop}%
\bibitem [{\citenamefont {Wang}\ \emph {et~al.}(2015)\citenamefont {Wang},
  \citenamefont {Ma},\ and\ \citenamefont {Ma}}]{wang2015}%
  \BibitemOpen
  \bibfield  {author} {\bibinfo {author} {\bibfnamefont {Z.-Y.}\ \bibnamefont
  {Wang}}, \bibinfo {author} {\bibfnamefont {Z.}~\bibnamefont {Ma}}, \ and\
  \bibinfo {author} {\bibfnamefont {Y.-q.}\ \bibnamefont {Ma}},\ }\href@noop {}
  {\bibfield  {journal} {\bibinfo  {journal} {Phys. Rev. E}\ }\textbf {\bibinfo
  {volume} {92}},\ \bibinfo {pages} {060303} (\bibinfo {year}
  {2015})}\BibitemShut {NoStop}%
\bibitem [{\citenamefont {Olvera de~la Cruz}\ \emph {et~al.}(1986)\citenamefont
  {Olvera de~la Cruz}, \citenamefont {Deutsch},\ and\ \citenamefont
  {Edwards}}]{de1986}%
  \BibitemOpen
  \bibfield  {author} {\bibinfo {author} {\bibfnamefont {M.}~\bibnamefont
  {Olvera de~la Cruz}}, \bibinfo {author} {\bibfnamefont {J.~M.}\ \bibnamefont
  {Deutsch}}, \ and\ \bibinfo {author} {\bibfnamefont {S.~F.}\ \bibnamefont
  {Edwards}},\ }\href@noop {} {\bibfield  {journal} {\bibinfo  {journal} {Phys.
  Rev. A}\ }\textbf {\bibinfo {volume} {33}},\ \bibinfo {pages} {2047}
  (\bibinfo {year} {1986})}\BibitemShut {NoStop}%
\bibitem [{\citenamefont {Shaffer}\ and\ \citenamefont {Olvera de~la
  Cruz}(1989)}]{shaffer1989}%
  \BibitemOpen
  \bibfield  {author} {\bibinfo {author} {\bibfnamefont {E.~O.}\ \bibnamefont
  {Shaffer}}\ and\ \bibinfo {author} {\bibfnamefont {M.}~\bibnamefont {Olvera
  de~la Cruz}},\ }\href@noop {} {\bibfield  {journal} {\bibinfo  {journal}
  {Macromolecules}\ }\textbf {\bibinfo {volume} {22}},\ \bibinfo {pages} {1351}
  (\bibinfo {year} {1989})}\BibitemShut {NoStop}%
\bibitem [{\citenamefont {Olvera de~la Cruz}\ \emph {et~al.}(1990)\citenamefont
  {Olvera de~la Cruz}, \citenamefont {Gersappe},\ and\ \citenamefont
  {Shaffer}}]{de1990}%
  \BibitemOpen
  \bibfield  {author} {\bibinfo {author} {\bibfnamefont {M.}~\bibnamefont
  {Olvera de~la Cruz}}, \bibinfo {author} {\bibfnamefont {D.}~\bibnamefont
  {Gersappe}}, \ and\ \bibinfo {author} {\bibfnamefont {E.~O.}\ \bibnamefont
  {Shaffer}},\ }\href@noop {} {\bibfield  {journal} {\bibinfo  {journal}
  {Physical review letters}\ }\textbf {\bibinfo {volume} {64}},\ \bibinfo
  {pages} {2324} (\bibinfo {year} {1990})}\BibitemShut {NoStop}%
\bibitem [{\citenamefont {Kaper}\ \emph {et~al.}(2003)\citenamefont {Kaper},
  \citenamefont {Busscher},\ and\ \citenamefont {Norde}}]{kaper2003}%
  \BibitemOpen
  \bibfield  {author} {\bibinfo {author} {\bibfnamefont {H.~J.}\ \bibnamefont
  {Kaper}}, \bibinfo {author} {\bibfnamefont {H.~J.}\ \bibnamefont {Busscher}},
  \ and\ \bibinfo {author} {\bibfnamefont {W.}~\bibnamefont {Norde}},\
  }\href@noop {} {\bibfield  {journal} {\bibinfo  {journal} {J. Biomater. Sci.
  Polym. Ed.}\ }\textbf {\bibinfo {volume} {14}},\ \bibinfo {pages} {313}
  (\bibinfo {year} {2003})}\BibitemShut {NoStop}%
\bibitem [{\citenamefont {Danger}\ \emph {et~al.}(2007)\citenamefont {Danger},
  \citenamefont {Ramonda},\ and\ \citenamefont {Cottet}}]{danger2007}%
  \BibitemOpen
  \bibfield  {author} {\bibinfo {author} {\bibfnamefont {G.}~\bibnamefont
  {Danger}}, \bibinfo {author} {\bibfnamefont {M.}~\bibnamefont {Ramonda}}, \
  and\ \bibinfo {author} {\bibfnamefont {H.}~\bibnamefont {Cottet}},\
  }\href@noop {} {\bibfield  {journal} {\bibinfo  {journal} {Electrophoresis}\
  }\textbf {\bibinfo {volume} {28}},\ \bibinfo {pages} {925} (\bibinfo {year}
  {2007})}\BibitemShut {NoStop}%
\bibitem [{\citenamefont {Nowack}\ and\ \citenamefont
  {Bucheli}(2007)}]{nowack2007}%
  \BibitemOpen
  \bibfield  {author} {\bibinfo {author} {\bibfnamefont {B.}~\bibnamefont
  {Nowack}}\ and\ \bibinfo {author} {\bibfnamefont {T.~D.}\ \bibnamefont
  {Bucheli}},\ }\href@noop {} {\bibfield  {journal} {\bibinfo  {journal}
  {Environ. Pollut.}\ }\textbf {\bibinfo {volume} {150}},\ \bibinfo {pages} {5}
  (\bibinfo {year} {2007})}\BibitemShut {NoStop}%
\bibitem [{\citenamefont {Obreg{\'o}n}\ \emph {et~al.}(2019)\citenamefont
  {Obreg{\'o}n}, \citenamefont {Amor},\ and\ \citenamefont
  {V{\'a}zquez}}]{obregon2019}%
  \BibitemOpen
  \bibfield  {author} {\bibinfo {author} {\bibfnamefont {S.}~\bibnamefont
  {Obreg{\'o}n}}, \bibinfo {author} {\bibfnamefont {G.}~\bibnamefont {Amor}}, \
  and\ \bibinfo {author} {\bibfnamefont {A.}~\bibnamefont {V{\'a}zquez}},\
  }\href@noop {} {\bibfield  {journal} {\bibinfo  {journal} {Adv. Colloid
  Interface Sci.}\ }\textbf {\bibinfo {volume} {269}},\ \bibinfo {pages} {236}
  (\bibinfo {year} {2019})}\BibitemShut {NoStop}%
\bibitem [{\citenamefont {von Smoluchowski}(1903)}]{von1903}%
  \BibitemOpen
  \bibfield  {author} {\bibinfo {author} {\bibfnamefont {M.}~\bibnamefont {von
  Smoluchowski}},\ }\href@noop {} {\bibfield  {journal} {\bibinfo  {journal}
  {Bull. Akad. Sci. Cracovie.}\ }\textbf {\bibinfo {volume} {8}},\ \bibinfo
  {pages} {182} (\bibinfo {year} {1903})}\BibitemShut {NoStop}%
\bibitem [{\citenamefont {Netz}\ and\ \citenamefont {Orland}(2000)}]{netz2000}%
  \BibitemOpen
  \bibfield  {author} {\bibinfo {author} {\bibfnamefont {R.~R.}\ \bibnamefont
  {Netz}}\ and\ \bibinfo {author} {\bibfnamefont {H.}~\bibnamefont {Orland}},\
  }\href@noop {} {\bibfield  {journal} {\bibinfo  {journal} {Eur. Phys. J. E}\
  }\textbf {\bibinfo {volume} {1}},\ \bibinfo {pages} {67} (\bibinfo {year}
  {2000})}\BibitemShut {NoStop}%
\bibitem [{\citenamefont {Chu}\ \emph {et~al.}(2007)\citenamefont {Chu},
  \citenamefont {Bai}, \citenamefont {Lipfert}, \citenamefont {Herschlag},\
  and\ \citenamefont {Doniach}}]{chu2007}%
  \BibitemOpen
  \bibfield  {author} {\bibinfo {author} {\bibfnamefont {V.~B.}\ \bibnamefont
  {Chu}}, \bibinfo {author} {\bibfnamefont {Y.}~\bibnamefont {Bai}}, \bibinfo
  {author} {\bibfnamefont {J.}~\bibnamefont {Lipfert}}, \bibinfo {author}
  {\bibfnamefont {D.}~\bibnamefont {Herschlag}}, \ and\ \bibinfo {author}
  {\bibfnamefont {S.}~\bibnamefont {Doniach}},\ }\href@noop {} {\bibfield
  {journal} {\bibinfo  {journal} {Biophys. J.}\ }\textbf {\bibinfo {volume}
  {93}},\ \bibinfo {pages} {3202} (\bibinfo {year} {2007})}\BibitemShut
  {NoStop}%
\bibitem [{\citenamefont {Borukhov}\ \emph {et~al.}(1997)\citenamefont
  {Borukhov}, \citenamefont {Andelman},\ and\ \citenamefont
  {Orland}}]{borukhov1997}%
  \BibitemOpen
  \bibfield  {author} {\bibinfo {author} {\bibfnamefont {I.}~\bibnamefont
  {Borukhov}}, \bibinfo {author} {\bibfnamefont {D.}~\bibnamefont {Andelman}},
  \ and\ \bibinfo {author} {\bibfnamefont {H.}~\bibnamefont {Orland}},\
  }\href@noop {} {\bibfield  {journal} {\bibinfo  {journal} {Phys. Rev. Lett.}\
  }\textbf {\bibinfo {volume} {79}},\ \bibinfo {pages} {435} (\bibinfo {year}
  {1997})}\BibitemShut {NoStop}%
\bibitem [{\citenamefont {Quesada-P{\'e}rez}\ \emph {et~al.}(2003)\citenamefont
  {Quesada-P{\'e}rez}, \citenamefont {Gonz{\'a}lez-Tovar}, \citenamefont
  {Mart{\'\i}n-Molina}, \citenamefont {Lozada-Cassou},\ and\ \citenamefont
  {Hidalgo-{\'A}lvarez}}]{quesada2003}%
  \BibitemOpen
  \bibfield  {author} {\bibinfo {author} {\bibfnamefont {M.}~\bibnamefont
  {Quesada-P{\'e}rez}}, \bibinfo {author} {\bibfnamefont {E.}~\bibnamefont
  {Gonz{\'a}lez-Tovar}}, \bibinfo {author} {\bibfnamefont {A.}~\bibnamefont
  {Mart{\'\i}n-Molina}}, \bibinfo {author} {\bibfnamefont {M.}~\bibnamefont
  {Lozada-Cassou}}, \ and\ \bibinfo {author} {\bibfnamefont {R.}~\bibnamefont
  {Hidalgo-{\'A}lvarez}},\ }\href@noop {} {\bibfield  {journal} {\bibinfo
  {journal} {ChemPhysChem}\ }\textbf {\bibinfo {volume} {4}},\ \bibinfo {pages}
  {234} (\bibinfo {year} {2003})}\BibitemShut {NoStop}%
\bibitem [{\citenamefont {Li}(2009)}]{li2009}%
  \BibitemOpen
  \bibfield  {author} {\bibinfo {author} {\bibfnamefont {B.}~\bibnamefont
  {Li}},\ }\href@noop {} {\bibfield  {journal} {\bibinfo  {journal}
  {Nonlinearity}\ }\textbf {\bibinfo {volume} {22}},\ \bibinfo {pages} {811}
  (\bibinfo {year} {2009})}\BibitemShut {NoStop}%
\bibitem [{\citenamefont {Gr{\o}nbech-Jensen}\ \emph
  {et~al.}(1997)\citenamefont {Gr{\o}nbech-Jensen}, \citenamefont {Mashl},
  \citenamefont {Bruinsma},\ and\ \citenamefont {Gelbart}}]{gronbech1997}%
  \BibitemOpen
  \bibfield  {author} {\bibinfo {author} {\bibfnamefont {N.}~\bibnamefont
  {Gr{\o}nbech-Jensen}}, \bibinfo {author} {\bibfnamefont {R.~J.}\ \bibnamefont
  {Mashl}}, \bibinfo {author} {\bibfnamefont {R.~F.}\ \bibnamefont {Bruinsma}},
  \ and\ \bibinfo {author} {\bibfnamefont {W.~M.}\ \bibnamefont {Gelbart}},\
  }\href@noop {} {\bibfield  {journal} {\bibinfo  {journal} {Phys. Rev. Lett.}\
  }\textbf {\bibinfo {volume} {78}},\ \bibinfo {pages} {2477} (\bibinfo {year}
  {1997})}\BibitemShut {NoStop}%
\bibitem [{\citenamefont {Rouzina}\ and\ \citenamefont
  {Bloomfield}(1996)}]{rouzina1996}%
  \BibitemOpen
  \bibfield  {author} {\bibinfo {author} {\bibfnamefont {I.}~\bibnamefont
  {Rouzina}}\ and\ \bibinfo {author} {\bibfnamefont {V.~A.}\ \bibnamefont
  {Bloomfield}},\ }\href@noop {} {\bibfield  {journal} {\bibinfo  {journal} {J.
  Phys. Chem.}\ }\textbf {\bibinfo {volume} {100}},\ \bibinfo {pages} {9977}
  (\bibinfo {year} {1996})}\BibitemShut {NoStop}%
\bibitem [{\citenamefont {Moreira}\ and\ \citenamefont
  {Netz}(2000)}]{moreira2000}%
  \BibitemOpen
  \bibfield  {author} {\bibinfo {author} {\bibfnamefont {A.~G.}\ \bibnamefont
  {Moreira}}\ and\ \bibinfo {author} {\bibfnamefont {R.~R.}\ \bibnamefont
  {Netz}},\ }\href@noop {} {\bibfield  {journal} {\bibinfo  {journal} {EPL}\
  }\textbf {\bibinfo {volume} {52}},\ \bibinfo {pages} {705} (\bibinfo {year}
  {2000})}\BibitemShut {NoStop}%
\bibitem [{\citenamefont {Netz}(2001)}]{netz2001}%
  \BibitemOpen
  \bibfield  {author} {\bibinfo {author} {\bibfnamefont {R.~R.}\ \bibnamefont
  {Netz}},\ }\href@noop {} {\bibfield  {journal} {\bibinfo  {journal} {Eur.
  Phys. J. E}\ }\textbf {\bibinfo {volume} {5}},\ \bibinfo {pages} {557}
  (\bibinfo {year} {2001})}\BibitemShut {NoStop}%
\bibitem [{\citenamefont {Netz}\ and\ \citenamefont {Orland}(2003)}]{netz2003}%
  \BibitemOpen
  \bibfield  {author} {\bibinfo {author} {\bibfnamefont {R.~R.}\ \bibnamefont
  {Netz}}\ and\ \bibinfo {author} {\bibfnamefont {H.}~\bibnamefont {Orland}},\
  }\href@noop {} {\bibfield  {journal} {\bibinfo  {journal} {Eur. Phys. J. E}\
  }\textbf {\bibinfo {volume} {11}},\ \bibinfo {pages} {301} (\bibinfo {year}
  {2003})}\BibitemShut {NoStop}%
\bibitem [{\citenamefont {Naji}\ \emph {et~al.}(2005)\citenamefont {Naji},
  \citenamefont {Jungblut}, \citenamefont {Moreira},\ and\ \citenamefont
  {Netz}}]{naji2005}%
  \BibitemOpen
  \bibfield  {author} {\bibinfo {author} {\bibfnamefont {A.}~\bibnamefont
  {Naji}}, \bibinfo {author} {\bibfnamefont {S.}~\bibnamefont {Jungblut}},
  \bibinfo {author} {\bibfnamefont {A.~G.}\ \bibnamefont {Moreira}}, \ and\
  \bibinfo {author} {\bibfnamefont {R.~R.}\ \bibnamefont {Netz}},\ }\href@noop
  {} {\bibfield  {journal} {\bibinfo  {journal} {Physica A}\ }\textbf {\bibinfo
  {volume} {352}},\ \bibinfo {pages} {131} (\bibinfo {year}
  {2005})}\BibitemShut {NoStop}%
\bibitem [{\citenamefont {Hatlo}\ and\ \citenamefont {Lue}(2010)}]{hatlo2010}%
  \BibitemOpen
  \bibfield  {author} {\bibinfo {author} {\bibfnamefont {M.~M.}\ \bibnamefont
  {Hatlo}}\ and\ \bibinfo {author} {\bibfnamefont {L.}~\bibnamefont {Lue}},\
  }\href@noop {} {\bibfield  {journal} {\bibinfo  {journal} {EPL}\ }\textbf
  {\bibinfo {volume} {89}},\ \bibinfo {pages} {25002} (\bibinfo {year}
  {2010})}\BibitemShut {NoStop}%
\bibitem [{\citenamefont {Buyukdagli}\ \emph {et~al.}(2012)\citenamefont
  {Buyukdagli}, \citenamefont {Achim},\ and\ \citenamefont
  {Ala-Nissila}}]{buyukdagli2012}%
  \BibitemOpen
  \bibfield  {author} {\bibinfo {author} {\bibfnamefont {S.}~\bibnamefont
  {Buyukdagli}}, \bibinfo {author} {\bibfnamefont {C.}~\bibnamefont {Achim}}, \
  and\ \bibinfo {author} {\bibfnamefont {T.}~\bibnamefont {Ala-Nissila}},\
  }\href@noop {} {\bibfield  {journal} {\bibinfo  {journal} {J. Chem. Phys.}\
  }\textbf {\bibinfo {volume} {137}},\ \bibinfo {pages} {104902} (\bibinfo
  {year} {2012})}\BibitemShut {NoStop}%
\bibitem [{\citenamefont {Buyukdagli}\ and\ \citenamefont
  {Ala-Nissila}(2014{\natexlab{a}})}]{buyukdagli2014}%
  \BibitemOpen
  \bibfield  {author} {\bibinfo {author} {\bibfnamefont {S.}~\bibnamefont
  {Buyukdagli}}\ and\ \bibinfo {author} {\bibfnamefont {T.}~\bibnamefont
  {Ala-Nissila}},\ }\href@noop {} {\bibfield  {journal} {\bibinfo  {journal}
  {Langmuir}\ }\textbf {\bibinfo {volume} {30}},\ \bibinfo {pages} {12907}
  (\bibinfo {year} {2014}{\natexlab{a}})}\BibitemShut {NoStop}%
\bibitem [{\citenamefont {Buyukdagli}\ and\ \citenamefont
  {Ala-Nissila}(2014{\natexlab{b}})}]{buyukdagli2014-2}%
  \BibitemOpen
  \bibfield  {author} {\bibinfo {author} {\bibfnamefont {S.}~\bibnamefont
  {Buyukdagli}}\ and\ \bibinfo {author} {\bibfnamefont {T.}~\bibnamefont
  {Ala-Nissila}},\ }\href@noop {} {\bibfield  {journal} {\bibinfo  {journal}
  {J. Chem. Phys.}\ }\textbf {\bibinfo {volume} {140}},\ \bibinfo {pages}
  {064701} (\bibinfo {year} {2014}{\natexlab{b}})}\BibitemShut {NoStop}%
\bibitem [{\citenamefont {Buyukdagli}\ and\ \citenamefont
  {Ala-Nissila}(2017)}]{buyukdagli2017}%
  \BibitemOpen
  \bibfield  {author} {\bibinfo {author} {\bibfnamefont {S.}~\bibnamefont
  {Buyukdagli}}\ and\ \bibinfo {author} {\bibfnamefont {T.}~\bibnamefont
  {Ala-Nissila}},\ }\href@noop {} {\bibfield  {journal} {\bibinfo  {journal}
  {J. Chem. Phys.}\ }\textbf {\bibinfo {volume} {147}},\ \bibinfo {pages}
  {144901} (\bibinfo {year} {2017})}\BibitemShut {NoStop}%
\bibitem [{\citenamefont {Gonz{\'a}lez-Tovar}\ \emph
  {et~al.}(2018)\citenamefont {Gonz{\'a}lez-Tovar}, \citenamefont
  {Lozada-Cassou}, \citenamefont {Bhuiyan},\ and\ \citenamefont
  {Outhwaite}}]{gonzalez2018}%
  \BibitemOpen
  \bibfield  {author} {\bibinfo {author} {\bibfnamefont {E.}~\bibnamefont
  {Gonz{\'a}lez-Tovar}}, \bibinfo {author} {\bibfnamefont {M.}~\bibnamefont
  {Lozada-Cassou}}, \bibinfo {author} {\bibfnamefont {L.~B.}\ \bibnamefont
  {Bhuiyan}}, \ and\ \bibinfo {author} {\bibfnamefont {C.~W.}\ \bibnamefont
  {Outhwaite}},\ }\href@noop {} {\bibfield  {journal} {\bibinfo  {journal} {J.
  Mol. Liq.}\ }\textbf {\bibinfo {volume} {270}},\ \bibinfo {pages} {157}
  (\bibinfo {year} {2018})}\BibitemShut {NoStop}%
\bibitem [{\citenamefont {Cats}\ \emph {et~al.}(2021)\citenamefont {Cats},
  \citenamefont {Evans}, \citenamefont {H{\"a}rtel},\ and\ \citenamefont
  {Van~Roij}}]{cats2021}%
  \BibitemOpen
  \bibfield  {author} {\bibinfo {author} {\bibfnamefont {P.}~\bibnamefont
  {Cats}}, \bibinfo {author} {\bibfnamefont {R.}~\bibnamefont {Evans}},
  \bibinfo {author} {\bibfnamefont {A.}~\bibnamefont {H{\"a}rtel}}, \ and\
  \bibinfo {author} {\bibfnamefont {R.}~\bibnamefont {Van~Roij}},\ }\href@noop
  {} {\bibfield  {journal} {\bibinfo  {journal} {J. Chem. Phys.}\ }\textbf
  {\bibinfo {volume} {154}},\ \bibinfo {pages} {124504} (\bibinfo {year}
  {2021})}\BibitemShut {NoStop}%
\bibitem [{\citenamefont {Cats}\ \emph {et~al.}(2022)\citenamefont {Cats},
  \citenamefont {Sitlapersad}, \citenamefont {den Otter}, \citenamefont
  {Thornton},\ and\ \citenamefont {Van~Roij}}]{cats2022}%
  \BibitemOpen
  \bibfield  {author} {\bibinfo {author} {\bibfnamefont {P.}~\bibnamefont
  {Cats}}, \bibinfo {author} {\bibfnamefont {R.~S.}\ \bibnamefont
  {Sitlapersad}}, \bibinfo {author} {\bibfnamefont {W.~K.}\ \bibnamefont {den
  Otter}}, \bibinfo {author} {\bibfnamefont {A.~R.}\ \bibnamefont {Thornton}},
  \ and\ \bibinfo {author} {\bibfnamefont {R.}~\bibnamefont {Van~Roij}},\
  }\href@noop {} {\bibfield  {journal} {\bibinfo  {journal} {J. Solution
  Chem.}\ }\textbf {\bibinfo {volume} {51}},\ \bibinfo {pages} {296} (\bibinfo
  {year} {2022})}\BibitemShut {NoStop}%
\bibitem [{\citenamefont {B{\"u}ltmann}\ and\ \citenamefont
  {H{\"a}rtel}(2022)}]{bultmann2022}%
  \BibitemOpen
  \bibfield  {author} {\bibinfo {author} {\bibfnamefont {M.}~\bibnamefont
  {B{\"u}ltmann}}\ and\ \bibinfo {author} {\bibfnamefont {A.}~\bibnamefont
  {H{\"a}rtel}},\ }\href@noop {} {\bibfield  {journal} {\bibinfo  {journal} {J.
  Phys. Condens. Matter}\ }\textbf {\bibinfo {volume} {34}},\ \bibinfo {pages}
  {235101} (\bibinfo {year} {2022})}\BibitemShut {NoStop}%
\bibitem [{\citenamefont {Deserno}\ \emph {et~al.}(2000)\citenamefont
  {Deserno}, \citenamefont {Holm},\ and\ \citenamefont {May}}]{deserno2000}%
  \BibitemOpen
  \bibfield  {author} {\bibinfo {author} {\bibfnamefont {M.}~\bibnamefont
  {Deserno}}, \bibinfo {author} {\bibfnamefont {C.}~\bibnamefont {Holm}}, \
  and\ \bibinfo {author} {\bibfnamefont {S.}~\bibnamefont {May}},\ }\href@noop
  {} {\bibfield  {journal} {\bibinfo  {journal} {Macromolecules}\ }\textbf
  {\bibinfo {volume} {33}},\ \bibinfo {pages} {199} (\bibinfo {year}
  {2000})}\BibitemShut {NoStop}%
\bibitem [{\citenamefont {Bagchi}\ and\ \citenamefont {Olvera de~la
  Cruz}(2020)}]{bagchi2020}%
  \BibitemOpen
  \bibfield  {author} {\bibinfo {author} {\bibfnamefont {D.}~\bibnamefont
  {Bagchi}}\ and\ \bibinfo {author} {\bibfnamefont {M.}~\bibnamefont {Olvera
  de~la Cruz}},\ }\href@noop {} {\bibfield  {journal} {\bibinfo  {journal} {J.
  Chem. Phys.}\ }\textbf {\bibinfo {volume} {153}},\ \bibinfo {pages} {184904}
  (\bibinfo {year} {2020})}\BibitemShut {NoStop}%
\bibitem [{\citenamefont {Gelbart}\ \emph {et~al.}(2000)\citenamefont
  {Gelbart}, \citenamefont {Bruinsma}, \citenamefont {Pincus},\ and\
  \citenamefont {Parsegian}}]{gelbart2000}%
  \BibitemOpen
  \bibfield  {author} {\bibinfo {author} {\bibfnamefont {W.~M.}\ \bibnamefont
  {Gelbart}}, \bibinfo {author} {\bibfnamefont {R.~F.}\ \bibnamefont
  {Bruinsma}}, \bibinfo {author} {\bibfnamefont {P.~A.}\ \bibnamefont
  {Pincus}}, \ and\ \bibinfo {author} {\bibfnamefont {V.~A.}\ \bibnamefont
  {Parsegian}},\ }\href@noop {} {\bibfield  {journal} {\bibinfo  {journal}
  {Phys. Today}\ }\textbf {\bibinfo {volume} {53}},\ \bibinfo {pages} {38}
  (\bibinfo {year} {2000})}\BibitemShut {NoStop}%
\bibitem [{\citenamefont {Tang}\ \emph {et~al.}(1997)\citenamefont {Tang},
  \citenamefont {Ito}, \citenamefont {Tao}, \citenamefont {Traub},\ and\
  \citenamefont {Janmey}}]{tang1997}%
  \BibitemOpen
  \bibfield  {author} {\bibinfo {author} {\bibfnamefont {J.~X.}\ \bibnamefont
  {Tang}}, \bibinfo {author} {\bibfnamefont {T.}~\bibnamefont {Ito}}, \bibinfo
  {author} {\bibfnamefont {T.}~\bibnamefont {Tao}}, \bibinfo {author}
  {\bibfnamefont {P.}~\bibnamefont {Traub}}, \ and\ \bibinfo {author}
  {\bibfnamefont {P.~A.}\ \bibnamefont {Janmey}},\ }\href@noop {} {\bibfield
  {journal} {\bibinfo  {journal} {Biochemistry}\ }\textbf {\bibinfo {volume}
  {36}},\ \bibinfo {pages} {12600} (\bibinfo {year} {1997})}\BibitemShut
  {NoStop}%
\bibitem [{\citenamefont {Angelini}\ \emph {et~al.}(2003)\citenamefont
  {Angelini}, \citenamefont {Liang}, \citenamefont {Wriggers},\ and\
  \citenamefont {Wong}}]{angelini2003}%
  \BibitemOpen
  \bibfield  {author} {\bibinfo {author} {\bibfnamefont {T.~E.}\ \bibnamefont
  {Angelini}}, \bibinfo {author} {\bibfnamefont {H.}~\bibnamefont {Liang}},
  \bibinfo {author} {\bibfnamefont {W.}~\bibnamefont {Wriggers}}, \ and\
  \bibinfo {author} {\bibfnamefont {G.~C.~L.}\ \bibnamefont {Wong}},\
  }\href@noop {} {\bibfield  {journal} {\bibinfo  {journal} {Proc. Natl. Acad.
  Sci.}\ }\textbf {\bibinfo {volume} {100}},\ \bibinfo {pages} {8634} (\bibinfo
  {year} {2003})}\BibitemShut {NoStop}%
\bibitem [{\citenamefont {Bernal}\ and\ \citenamefont
  {Fankuchen}(1941)}]{bernal1941}%
  \BibitemOpen
  \bibfield  {author} {\bibinfo {author} {\bibfnamefont {J.~D.}\ \bibnamefont
  {Bernal}}\ and\ \bibinfo {author} {\bibfnamefont {I.}~\bibnamefont
  {Fankuchen}},\ }\href@noop {} {\bibfield  {journal} {\bibinfo  {journal} {J.
  Gen. Physiol.}\ }\textbf {\bibinfo {volume} {25}},\ \bibinfo {pages} {111}
  (\bibinfo {year} {1941})}\BibitemShut {NoStop}%
\bibitem [{\citenamefont {Kominami}\ \emph {et~al.}(2019)\citenamefont
  {Kominami}, \citenamefont {Kobayashi},\ and\ \citenamefont
  {Yamada}}]{kominami2019molecular}%
  \BibitemOpen
  \bibfield  {author} {\bibinfo {author} {\bibfnamefont {H.}~\bibnamefont
  {Kominami}}, \bibinfo {author} {\bibfnamefont {K.}~\bibnamefont {Kobayashi}},
  \ and\ \bibinfo {author} {\bibfnamefont {H.}~\bibnamefont {Yamada}},\
  }\href@noop {} {\bibfield  {journal} {\bibinfo  {journal} {Sci. Rep.}\
  }\textbf {\bibinfo {volume} {9}},\ \bibinfo {pages} {6851} (\bibinfo {year}
  {2019})}\BibitemShut {NoStop}%
\bibitem [{\citenamefont {Weeks}\ \emph {et~al.}(1971)\citenamefont {Weeks},
  \citenamefont {Chandler},\ and\ \citenamefont {Andersen}}]{anderson}%
  \BibitemOpen
  \bibfield  {author} {\bibinfo {author} {\bibfnamefont {J.~D.}\ \bibnamefont
  {Weeks}}, \bibinfo {author} {\bibfnamefont {D.}~\bibnamefont {Chandler}}, \
  and\ \bibinfo {author} {\bibfnamefont {H.~C.}\ \bibnamefont {Andersen}},\
  }\href@noop {} {\bibfield  {journal} {\bibinfo  {journal} {J. Chem. Phys.}\
  }\textbf {\bibinfo {volume} {54}},\ \bibinfo {pages} {5237} (\bibinfo {year}
  {1971})}\BibitemShut {NoStop}%
\bibitem [{\citenamefont {Plimpton}\ \emph {et~al.}(1997)\citenamefont
  {Plimpton}, \citenamefont {Pollock},\ and\ \citenamefont
  {Stevens}}]{plimpton1997}%
  \BibitemOpen
  \bibfield  {author} {\bibinfo {author} {\bibfnamefont {S.}~\bibnamefont
  {Plimpton}}, \bibinfo {author} {\bibfnamefont {R.}~\bibnamefont {Pollock}}, \
  and\ \bibinfo {author} {\bibfnamefont {M.}~\bibnamefont {Stevens}},\
  }\href@noop {} {\emph {\bibinfo {title} {in Proceedings of the Eighth SIAM
  Conference on Parallel Processing for Scientific Computing}}}\ (\bibinfo
  {publisher} {SIAM},\ \bibinfo {address} {Minneapolis},\ \bibinfo {year}
  {1997})\BibitemShut {NoStop}%
\bibitem [{\citenamefont {Plimpton}(1995)}]{plimpton1995}%
  \BibitemOpen
  \bibfield  {author} {\bibinfo {author} {\bibfnamefont {S.}~\bibnamefont
  {Plimpton}},\ }\href@noop {} {\bibfield  {journal} {\bibinfo  {journal} {J.
  Comput. Phys.}\ }\textbf {\bibinfo {volume} {117}},\ \bibinfo {pages} {1}
  (\bibinfo {year} {1995})}\BibitemShut {NoStop}%
\bibitem [{\citenamefont {Brown}\ \emph {et~al.}(2009)\citenamefont {Brown},
  \citenamefont {Petersen}, \citenamefont {Plimpton},\ and\ \citenamefont
  {Grest}}]{brown2009}%
  \BibitemOpen
  \bibfield  {author} {\bibinfo {author} {\bibfnamefont {W.~M.}\ \bibnamefont
  {Brown}}, \bibinfo {author} {\bibfnamefont {M.~K.}\ \bibnamefont {Petersen}},
  \bibinfo {author} {\bibfnamefont {S.~J.}\ \bibnamefont {Plimpton}}, \ and\
  \bibinfo {author} {\bibfnamefont {G.~S.}\ \bibnamefont {Grest}},\ }\href@noop
  {} {\bibfield  {journal} {\bibinfo  {journal} {J. Chem. Phys.}\ }\textbf
  {\bibinfo {volume} {130}},\ \bibinfo {pages} {044901} (\bibinfo {year}
  {2009})}\BibitemShut {NoStop}%
\bibitem [{\citenamefont {Thompson}\ \emph {et~al.}(2022)\citenamefont
  {Thompson}, \citenamefont {Aktulga}, \citenamefont {Berger}, \citenamefont
  {Bolintineanu}, \citenamefont {Brown}, \citenamefont {Crozier}, \citenamefont
  {Veld}, \citenamefont {Kohlmeyer}, \citenamefont {Moore}, \citenamefont
  {Nguyen}, \citenamefont {Shan}, \citenamefont {Stevens}, \citenamefont
  {Tranchida}, \citenamefont {Trott},\ and\ \citenamefont
  {Plimpton}}]{Thomson2022}%
  \BibitemOpen
  \bibfield  {author} {\bibinfo {author} {\bibfnamefont {A.~P.}\ \bibnamefont
  {Thompson}}, \bibinfo {author} {\bibfnamefont {H.~M.}\ \bibnamefont
  {Aktulga}}, \bibinfo {author} {\bibfnamefont {R.}~\bibnamefont {Berger}},
  \bibinfo {author} {\bibfnamefont {D.~S.}\ \bibnamefont {Bolintineanu}},
  \bibinfo {author} {\bibfnamefont {W.~M.}\ \bibnamefont {Brown}}, \bibinfo
  {author} {\bibfnamefont {P.~S.}\ \bibnamefont {Crozier}}, \bibinfo {author}
  {\bibfnamefont {P.~I.}\ \bibnamefont {Veld}}, \bibinfo {author}
  {\bibfnamefont {A.}~\bibnamefont {Kohlmeyer}}, \bibinfo {author}
  {\bibfnamefont {S.~G.}\ \bibnamefont {Moore}}, \bibinfo {author}
  {\bibfnamefont {T.~D.}\ \bibnamefont {Nguyen}}, \bibinfo {author}
  {\bibfnamefont {R.}~\bibnamefont {Shan}}, \bibinfo {author} {\bibfnamefont
  {M.~J.}\ \bibnamefont {Stevens}}, \bibinfo {author} {\bibfnamefont
  {J.}~\bibnamefont {Tranchida}}, \bibinfo {author} {\bibfnamefont
  {C.}~\bibnamefont {Trott}}, \ and\ \bibinfo {author} {\bibfnamefont {S.~J.}\
  \bibnamefont {Plimpton}},\ }\href {\doibase 10.1016/j.cpc.2021.108171}
  {\bibfield  {journal} {\bibinfo  {journal} {Comp. Phys. Comm.}\ }\textbf
  {\bibinfo {volume} {271}},\ \bibinfo {pages} {108171} (\bibinfo {year}
  {2022})}\BibitemShut {NoStop}%
\bibitem [{\citenamefont {Nos{\'e}}(1984)}]{nose1984}%
  \BibitemOpen
  \bibfield  {author} {\bibinfo {author} {\bibfnamefont {S.}~\bibnamefont
  {Nos{\'e}}},\ }\href@noop {} {\bibfield  {journal} {\bibinfo  {journal} {Mol.
  Phys.}\ }\textbf {\bibinfo {volume} {52}},\ \bibinfo {pages} {255} (\bibinfo
  {year} {1984})}\BibitemShut {NoStop}%
\bibitem [{\citenamefont {Hoover}(1985)}]{hoover1985}%
  \BibitemOpen
  \bibfield  {author} {\bibinfo {author} {\bibfnamefont {W.~G.}\ \bibnamefont
  {Hoover}},\ }\href@noop {} {\bibfield  {journal} {\bibinfo  {journal} {Phys.
  Rev. A}\ }\textbf {\bibinfo {volume} {31}},\ \bibinfo {pages} {1695}
  (\bibinfo {year} {1985})}\BibitemShut {NoStop}%
\bibitem [{\citenamefont {Jewett}\ \emph {et~al.}(2021)\citenamefont {Jewett},
  \citenamefont {Stelter}, \citenamefont {Lambert}, \citenamefont {Saladi},
  \citenamefont {Roscioni}, \citenamefont {Ricci}, \citenamefont {Autin},
  \citenamefont {Maritan}, \citenamefont {Bashusqeh}, \citenamefont {Keyes}
  \emph {et~al.}}]{jewett2021}%
  \BibitemOpen
  \bibfield  {author} {\bibinfo {author} {\bibfnamefont {A.~I.}\ \bibnamefont
  {Jewett}}, \bibinfo {author} {\bibfnamefont {D.}~\bibnamefont {Stelter}},
  \bibinfo {author} {\bibfnamefont {J.}~\bibnamefont {Lambert}}, \bibinfo
  {author} {\bibfnamefont {S.~M.}\ \bibnamefont {Saladi}}, \bibinfo {author}
  {\bibfnamefont {O.~M.}\ \bibnamefont {Roscioni}}, \bibinfo {author}
  {\bibfnamefont {M.}~\bibnamefont {Ricci}}, \bibinfo {author} {\bibfnamefont
  {L.}~\bibnamefont {Autin}}, \bibinfo {author} {\bibfnamefont
  {M.}~\bibnamefont {Maritan}}, \bibinfo {author} {\bibfnamefont {S.~M.}\
  \bibnamefont {Bashusqeh}}, \bibinfo {author} {\bibfnamefont {T.}~\bibnamefont
  {Keyes}},  \emph {et~al.},\ }\href@noop {} {\bibfield  {journal} {\bibinfo
  {journal} {J. Mol. Biol.}\ }\textbf {\bibinfo {volume} {433}},\ \bibinfo
  {pages} {166841} (\bibinfo {year} {2021})}\BibitemShut {NoStop}%
\bibitem [{\citenamefont {Joly}\ \emph {et~al.}(2006)\citenamefont {Joly},
  \citenamefont {Ybert}, \citenamefont {Trizac},\ and\ \citenamefont
  {Bocquet}}]{joly2006}%
  \BibitemOpen
  \bibfield  {author} {\bibinfo {author} {\bibfnamefont {L.}~\bibnamefont
  {Joly}}, \bibinfo {author} {\bibfnamefont {C.}~\bibnamefont {Ybert}},
  \bibinfo {author} {\bibfnamefont {E.}~\bibnamefont {Trizac}}, \ and\ \bibinfo
  {author} {\bibfnamefont {L.}~\bibnamefont {Bocquet}},\ }\href@noop {}
  {\bibfield  {journal} {\bibinfo  {journal} {J. Chem. Phys.}\ }\textbf
  {\bibinfo {volume} {125}},\ \bibinfo {pages} {204716} (\bibinfo {year}
  {2006})}\BibitemShut {NoStop}%
\bibitem [{\citenamefont {Galla}\ \emph {et~al.}(2014)\citenamefont {Galla},
  \citenamefont {Meyer}, \citenamefont {Spiering}, \citenamefont {Sischka},
  \citenamefont {Mayer}, \citenamefont {Hall}, \citenamefont {Reimann},\ and\
  \citenamefont {Anselmetti}}]{galla2014}%
  \BibitemOpen
  \bibfield  {author} {\bibinfo {author} {\bibfnamefont {L.}~\bibnamefont
  {Galla}}, \bibinfo {author} {\bibfnamefont {A.~J.}\ \bibnamefont {Meyer}},
  \bibinfo {author} {\bibfnamefont {A.}~\bibnamefont {Spiering}}, \bibinfo
  {author} {\bibfnamefont {A.}~\bibnamefont {Sischka}}, \bibinfo {author}
  {\bibfnamefont {M.}~\bibnamefont {Mayer}}, \bibinfo {author} {\bibfnamefont
  {A.~R.}\ \bibnamefont {Hall}}, \bibinfo {author} {\bibfnamefont
  {P.}~\bibnamefont {Reimann}}, \ and\ \bibinfo {author} {\bibfnamefont
  {D.}~\bibnamefont {Anselmetti}},\ }\href@noop {} {\bibfield  {journal}
  {\bibinfo  {journal} {Nano Lett.}\ }\textbf {\bibinfo {volume} {14}},\
  \bibinfo {pages} {4176} (\bibinfo {year} {2014})}\BibitemShut {NoStop}%
\bibitem [{\citenamefont {Raafatnia}\ \emph {et~al.}(2014)\citenamefont
  {Raafatnia}, \citenamefont {Hickey},\ and\ \citenamefont
  {Holm}}]{raafatnia2014}%
  \BibitemOpen
  \bibfield  {author} {\bibinfo {author} {\bibfnamefont {S.}~\bibnamefont
  {Raafatnia}}, \bibinfo {author} {\bibfnamefont {O.~A.}\ \bibnamefont
  {Hickey}}, \ and\ \bibinfo {author} {\bibfnamefont {C.}~\bibnamefont
  {Holm}},\ }\href@noop {} {\bibfield  {journal} {\bibinfo  {journal} {Phys.
  Rev. Lett.}\ }\textbf {\bibinfo {volume} {113}},\ \bibinfo {pages} {238301}
  (\bibinfo {year} {2014})}\BibitemShut {NoStop}%
\bibitem [{\citenamefont {Yamaguchi}\ and\ \citenamefont
  {Kobayashi}(2016)}]{yamaguchi2016}%
  \BibitemOpen
  \bibfield  {author} {\bibinfo {author} {\bibfnamefont {A.}~\bibnamefont
  {Yamaguchi}}\ and\ \bibinfo {author} {\bibfnamefont {M.}~\bibnamefont
  {Kobayashi}},\ }\href@noop {} {\bibfield  {journal} {\bibinfo  {journal}
  {Colloid Polym. Sci.}\ }\textbf {\bibinfo {volume} {294}},\ \bibinfo {pages}
  {1019} (\bibinfo {year} {2016})}\BibitemShut {NoStop}%
\end{thebibliography}%
\end{document}